\documentclass[pra]{revtex4}
\usepackage{graphicx}

\newcommand{\rr}{\mathbf{r}}
\newcommand{\kk}{\mathbf{k}}
\newcommand{\qq}{\mathbf{q}}
\newcommand{\vv}{\mathbf{v}}

\newcommand{\eqcite}[1]{(\ref{eq:#1})}
\newcommand{\eqname}[1]{\label{eq:#1}}

\begin{document}
\title[Superfluid light in propagating geometries]{Superfluid light in propagating geometries}
\author{Iacopo Carusotto}

\affiliation{INO-CNR BEC Center and Universit\`a di Trento, via Sommarive 14, I-38123 Povo, Italy}

\begin{abstract}
We review how the paraxial approximation naturally leads to a hydrodynamic description of light propagation  in a Kerr nonlinear medium analogous to the Gross-Pitaevskii equation for the temporal evolution of the order parameter of a superfluid.
The main features of the many-body collective dynamics of these fluids of light in a propagating geometry are discussed: Generation and observation of Bogoliubov sound waves on top of the fluid is first described. Experimentally accessible manifestations of superfluidity are then highlighted. Perspectives in view of realizing analog models of gravity are finally given.
\end{abstract}
\maketitle

\section{Introduction}

Experimental studies of the so-called fluids of light are opening new perspectives to the field of many-body physics, as they allow unprecedented control and flexibility in the generation, manipulation and control of Bose fluids~\cite{ICCCRMP}. So far, a number of striking experimental observation have been performed using a semiconductor planar microcavity architecture, including the demonstration of a superfluid flow~\cite{Amo2009} and of the hydrodynamic nucleation of solitons and quantized vortices~\cite{Amo11,Nardin,Sanvitto}. 

An alternative platform for studying many body physics in fluids of light consists of a bulk nonlinear crystal showing an intensity-dependent refractive index: under the paraxial approximation, the propagation of monochromatic light can be described in terms of a Gross-Pitaevskii equation for the order parameter, in our case the electric field amplitude of the monochromatic beam. Even though experimental studies of this system have started much earlier, up to now only a little attention has been devoted to hydrodynamic and superfluid features. Among the most remarkable exceptions, we may mention the recent works~\cite{Wan,Wan2,Jia,Wan3,Jia2,Khamis,Moulieras}.

This short article reports an application of superfluid hydrodynamics concepts to the theoretical study of classical light propagation in nonlinear crystals. Sec.\ref{sec:model} describes the system under consideration and gives a short derivation of the paraxial wave equation in a Kerr nonlinear medium: In contrast to the microcavity architecture where the dynamics of the fluid of light is described by a driven-dissipative equation, paraxial propagation is described by a fully conservative time-dependent Gross-Pitaevskii equation.
The generation and observation of elementary excitations on top of the fluid of light are reviewed in Sec.\ref{sec:phonon}. The interaction of a flowing fluid of light with a localized defect is discussed in Sec.\ref{sec:defect}: signatures of superfluid behavior are highlighted, as well as the main mechanisms for breaking superfluidity. Perspectives in view of using superfluids of light for experimental studies of analog models of gravity are finally outlined in Sec.\ref{sec:BH}. Conclusions are drawn in Sec.\ref{sec:conclu}.

\section{The model}
\label{sec:model}

\begin{figure}
\begin{center}
 \includegraphics[width=12cm,clip]{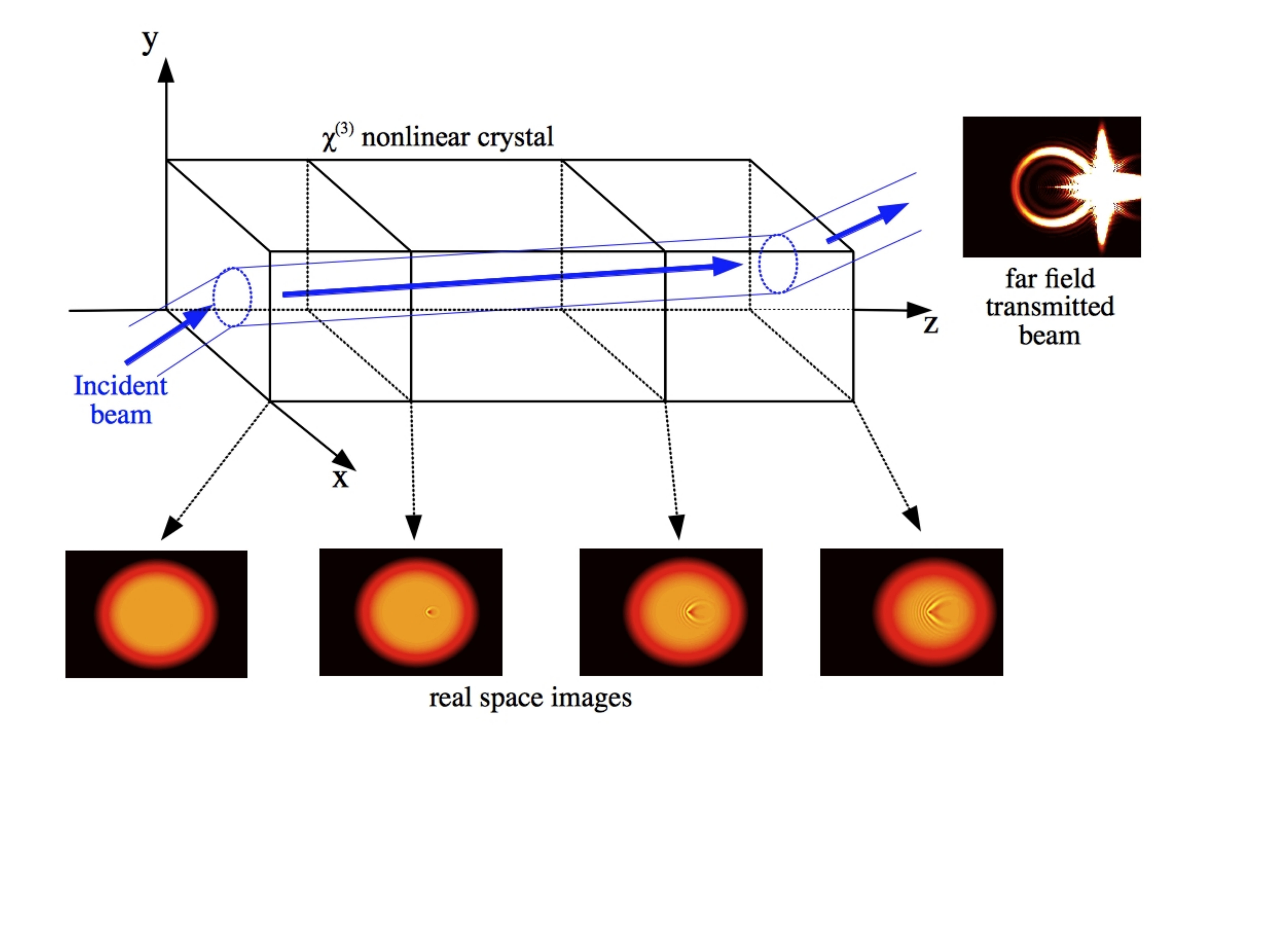}
 \vspace{0cm}
\caption{Sketch of the experimental configuration under investigation in the present work. The small panels in the bottom of the figure show an illustrative example of a series of intensity profiles after different distances of propagation along the crystal. The small panel on the right shows an example of far-field transmitted intensity profile.
\label{fig:sketch}}
\end{center}
 \end{figure}

The system we are considering is sketched in Fig.\ref{fig:sketch}: a monochromatic wave of frequency $\omega_0$ is incident on a medium of linear dielectric constant $\epsilon$ and Kerr optical nonlinearity $\chi^{(3)}$. The front interface of the crystal is assumed to lie on the $(x,y)$ plane, while the incident beam is assumed to propagate close to the longitudinal direction $z$.

Neglecting for simplicity polarization degrees of freedom, the propagation equation of light in the crystal can be described by the usual nonlinear wave equation
\begin{equation}
\partial_z^2 E(\rr_\perp,z) + \nabla_\perp^2 E(\rr_\perp,z)+ \frac{\omega_0^2}{c^2}\,\left[\epsilon+\delta \epsilon(\rr_\perp,z) + \chi^{(3)}\,|E(\rr_\perp,z)|^2\right]\, E(\rr_\perp,z) ,
\eqname{Maxw}
\end{equation}
for the complex amplitude $E(\rr_\perp,z)$ of the monochromatic field. The $\nabla_\perp$ gradient is taken along the transverse $\rr_\perp=(x,y)$ coordinates only. The spatial modulation of the linear dielectric constant $\delta \epsilon(\rr_\perp,z)$ is assumed to be slowly varying in space. A further slow modulation of the effective dieletric constant proportional to the local light intensity is provided by the following term involving the third-order Kerr optical nonlinearity $\chi^{(3)}$ of the medium.

Provided the variation of the field amplitude $E(\rr_\perp,z)$ along the transverse plane is slow enough, we can perform the so-called paraxial approximation. In terms of the slowly varying envelope $\mathcal{E}(\rr_\perp,z)=E(\rr_\perp,z)\,e^{-i k_0 z}$, the paraxial approximation is accurate as long as $ |\nabla_{\perp}^{2} \mathcal{E}|/ k_{0}^{2} \sim |\partial_{z} \mathcal{E}| / k_{0} \ll 1$ with $k_0=\sqrt{\epsilon} \,\omega_0 /c$. Under this assumption, the second $z$ derivative term can be neglected in the propagation equation for $\mathcal{E}$, which is then written in the form of a nonlinear Schr\"odinger equation,
\begin{equation}
i\partial_z \mathcal{E}(\rr_\perp,z)= -\frac{1}{2k_0} \nabla_\perp^2 \mathcal{E} - \frac{k_0}{2\epsilon} \left( \delta \epsilon(\rr_\perp,z) + \chi^{(3)}\, |\mathcal{E}(\rr_\perp,z)|^2 \right) \mathcal{E}(\rr_\perp,z).
\eqname{Schr}
\end{equation}
The effective mass is positive and proportional to $k_0$. The spatial modulation of the dielectric constant provides an external potential term $V(\rr_\perp,z)= -k_0 \delta\epsilon (\rr_\perp,z)/(2\epsilon)$: higher (lower) refractive index means attractive (repulsive) potential.
Analogously, the optical nonlinearity gives a photon-photon interaction constant $g=-\chi^{(3)}k_0/2\epsilon$: a focusing $\chi^{(3)}>0$ (defocusing $\chi^{(3)}<0$) optical nonlinearity corresponds to an attractive $g<0$ (repulsive $g>0$) effective interaction between photons. Throughout this work, we focus our attention on the more stable defocusing $\chi^{(3)}<0$ case giving repulsive $g>0$ interactions. We also restrict our attention to the case of a weak nonlinearity, where a large number of photons is collectively interacting to give the observed nonlinearity. 

In a typical experiment, a monochromatic laser beam is shone on the front surface of the crystal: the amplitude profile  of the incident beam $E_0(\rr_\perp)$ fixes the initial condition $\mathcal{E}(\rr_\perp,z=0)=E_0(\rr_\perp)$. The Schr\"odinger-like equation \eqcite{Schr} then describes how the transverse profile of the laser beam evolves during propagation along the nonlinear crystal. The interpretation of this optical phenomenon in terms of a {\em fluid of light} stems from the formal analogy of this equation with the Gross-Pitaevskii equation (GPE) for the order parameter of a superfluid or, equivalently, the macroscopic wave function of a dilute Bose-Einstein condensate~\cite{PSbook}. 

Before proceeding, it is however crucial to stress that while the standard GPE for superfluid Helium or ultracold atomic clouds describes the evolution of the macroscopic wavefunction in real time, our equation \eqcite{Schr} refers to a propagation in space: a space-time mapping is therefore understood when speaking about fluid of light in the present context. This seemingly minor issue will have profound consequences when one tries to build from \eqcite{Schr} a fully quantum field theory accounting also for the corpuscular nature of light. This task is of crucial importance to describe the strong nonlinearity case, where just a few photons are able to produce sizable nonlinear effects. First experimental studies of this novel regime have been recently reported using coherently a gas of dressed atoms in the so-called Rydberg-EIT regime~\cite{Lukin}.

\section{Sound waves in the fluid of light}
\label{sec:phonon}

Transposing to the present optical context the Bogoliubov theory of weak perturbations on top of a weakly interacting Bose gas~\cite{PSbook}, the dispersion of the elementary excitations on top of a spatially uniform fluid of light of density $|E_0|^2$ at rest has the form
\begin{equation}
W_{\rm Bog}(\kk_\perp)=\sqrt{\frac{k_\perp^2}{2k_0}\left(\frac{k_\perp^2}{2k_0} -\frac{k_0 \chi^{(3)}\, |E_0|^2}{\epsilon} \right)  }.
\end{equation}
Depending on the relative value of the excitation wavevector $k$ as compared to the so-called healing length $\xi=[-2\epsilon/(\chi^{(3)}\,|E_0|^2)]^{1/2}\,k_0^{-1}$, two regimes can be identified. Large momentum excitations with $k_\perp \xi\gg1$ have a single-particle character and a parabolic dispersion $W_{\rm Bog}\simeq k_\perp^2/(2k_0)$. Small momentum excitations with $k_\perp \xi \ll 1$ have a sonic dispersion $W_{\rm Bog}\simeq c_s k$ with a speed of sound 
\begin{equation}
  c_s=\left[-\frac{\chi^{(3)}\,|E_0|^2}{2\epsilon} \right]^{1/2}.
  \eqname{cs}
\end{equation}
Of course, this result only holds in the case of a defocusing $\chi^{(3)}<0$ nonlinearity when the photon-photon interaction is repulsive. In the opposite $\chi^{(3)}>0$ case, the attractive photon-photon interaction would make the fluid of light unstable against modulational instabilities, which is signalled by the Bogoliubov dispersion $W_{\rm Bog}(\kk_\perp)$ becoming imaginary at low wavevectors. In the optical language, this instability for a focusing nonlinearity $\chi^{(3)}>0$ goes under the name of filamentation of the laser beam. 

It is worth noting that, as a consequence of the space-time $t\leftrightarrow z$ mapping underlying the Schr\"odinger equation \eqcite{Schr}, frequencies $W_{\rm Bog}(\kk_\perp)$ are measured in inverse lengths and speeds like $c_s$ and $v_{\rm gr}=\nabla_\kk W_{\rm Bog}(\kk_\perp)$ are measured in adimensional units. For convenience, the wavelength $\lambda_0=2\pi/k_0$ will be used as a unit of length in all figures.

\begin{figure}[htbp]
\includegraphics[width=4cm]{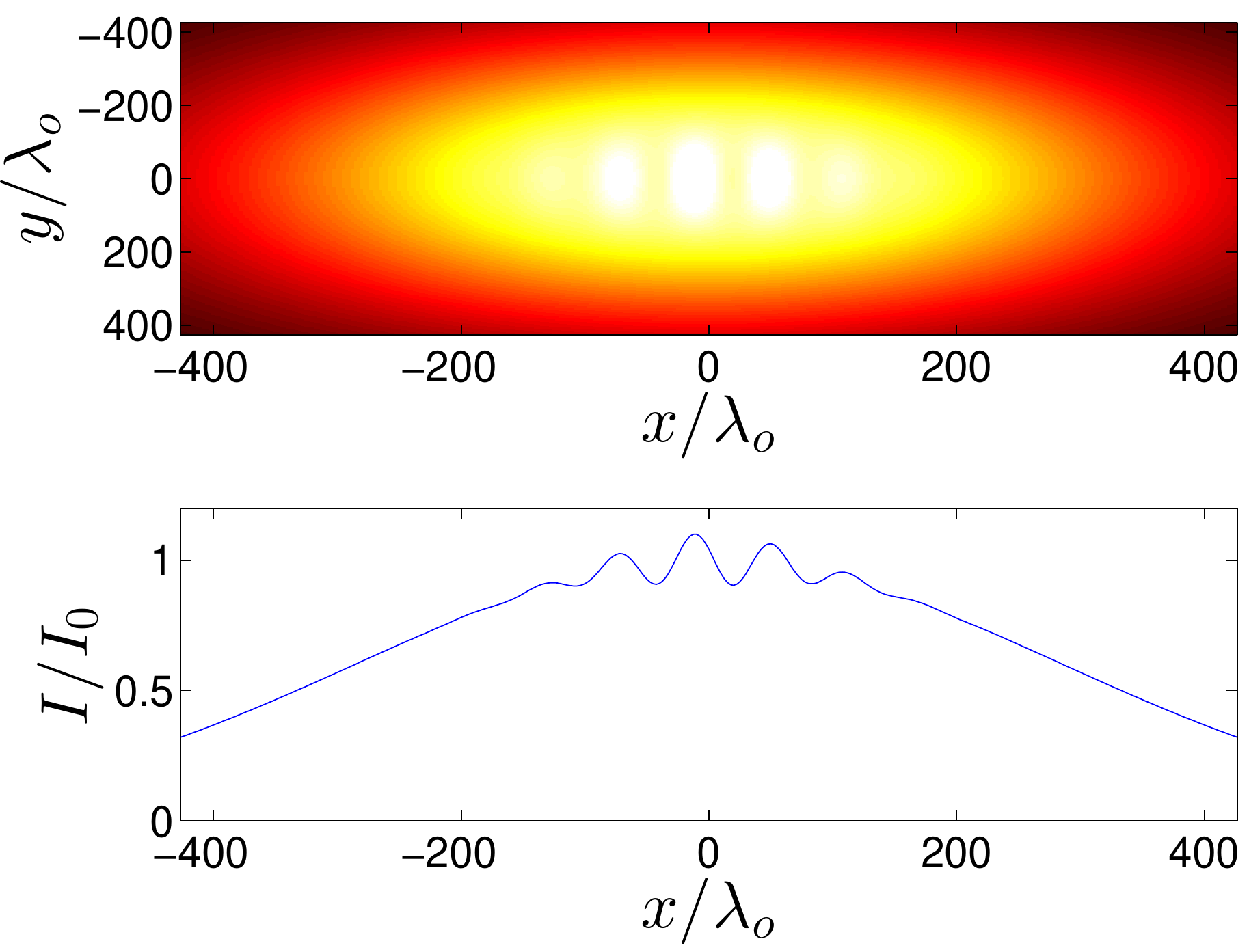}
\includegraphics[width=4cm]{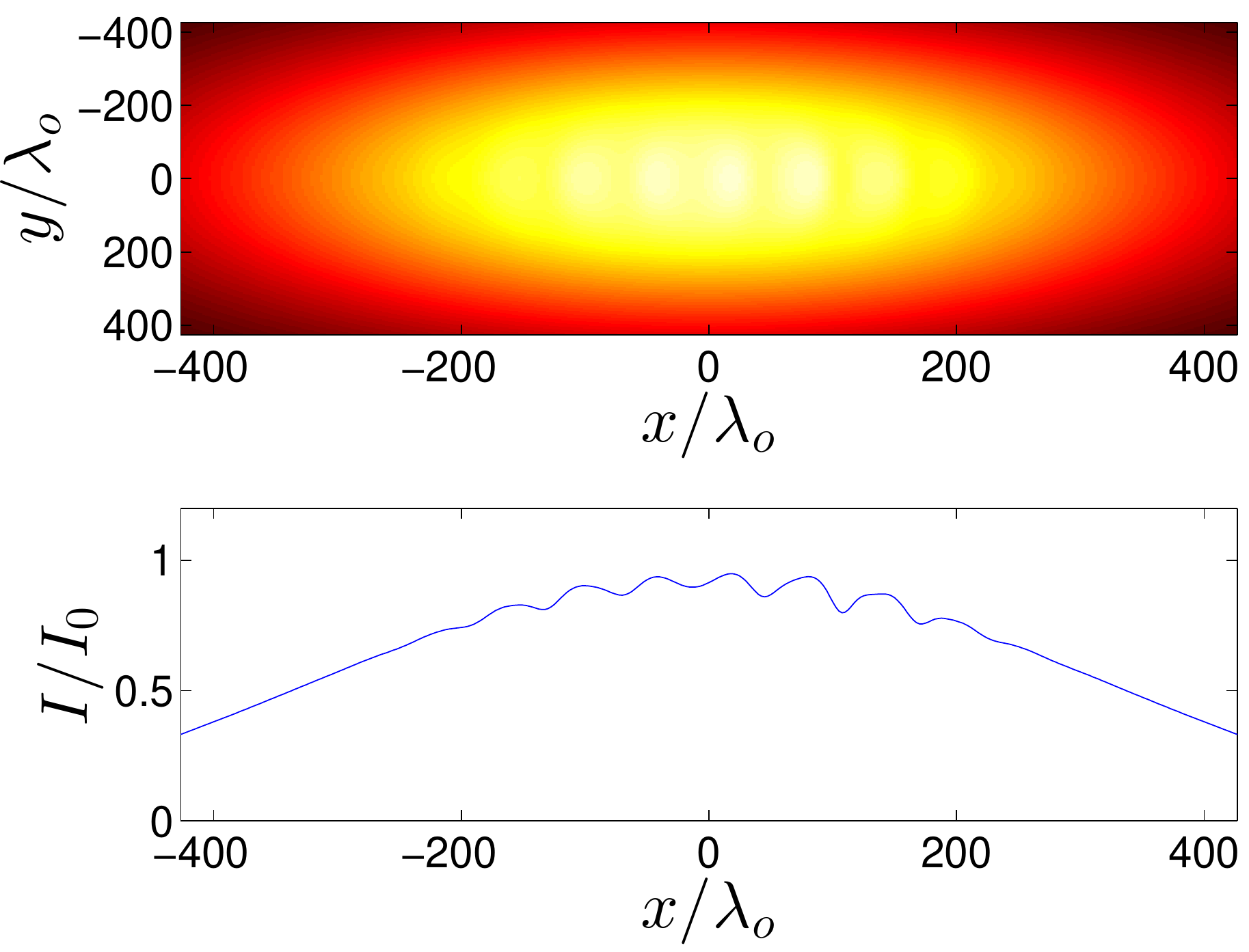}
\includegraphics[width=4cm]{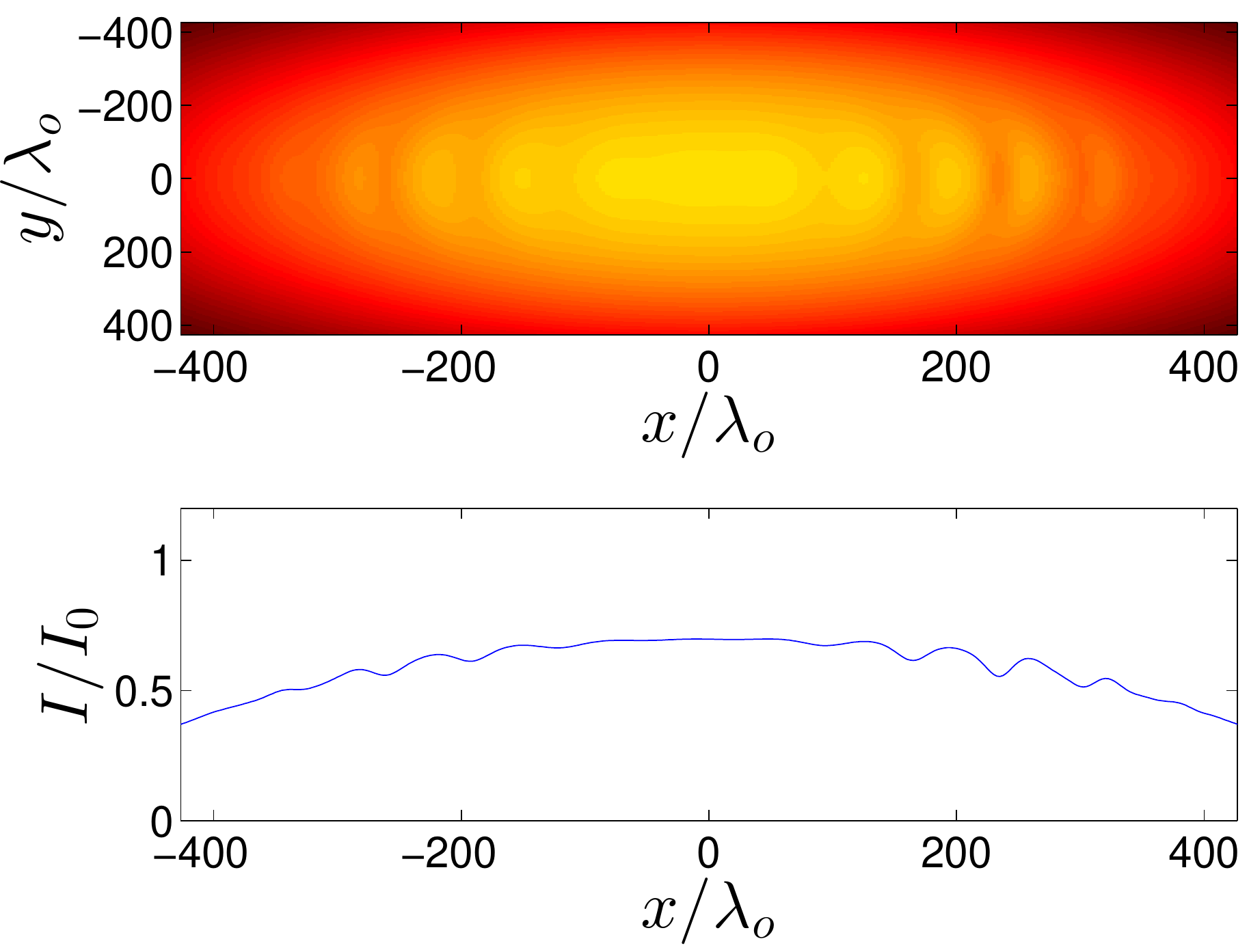}

\vspace*{0.5cm}
\includegraphics[width=4cm]{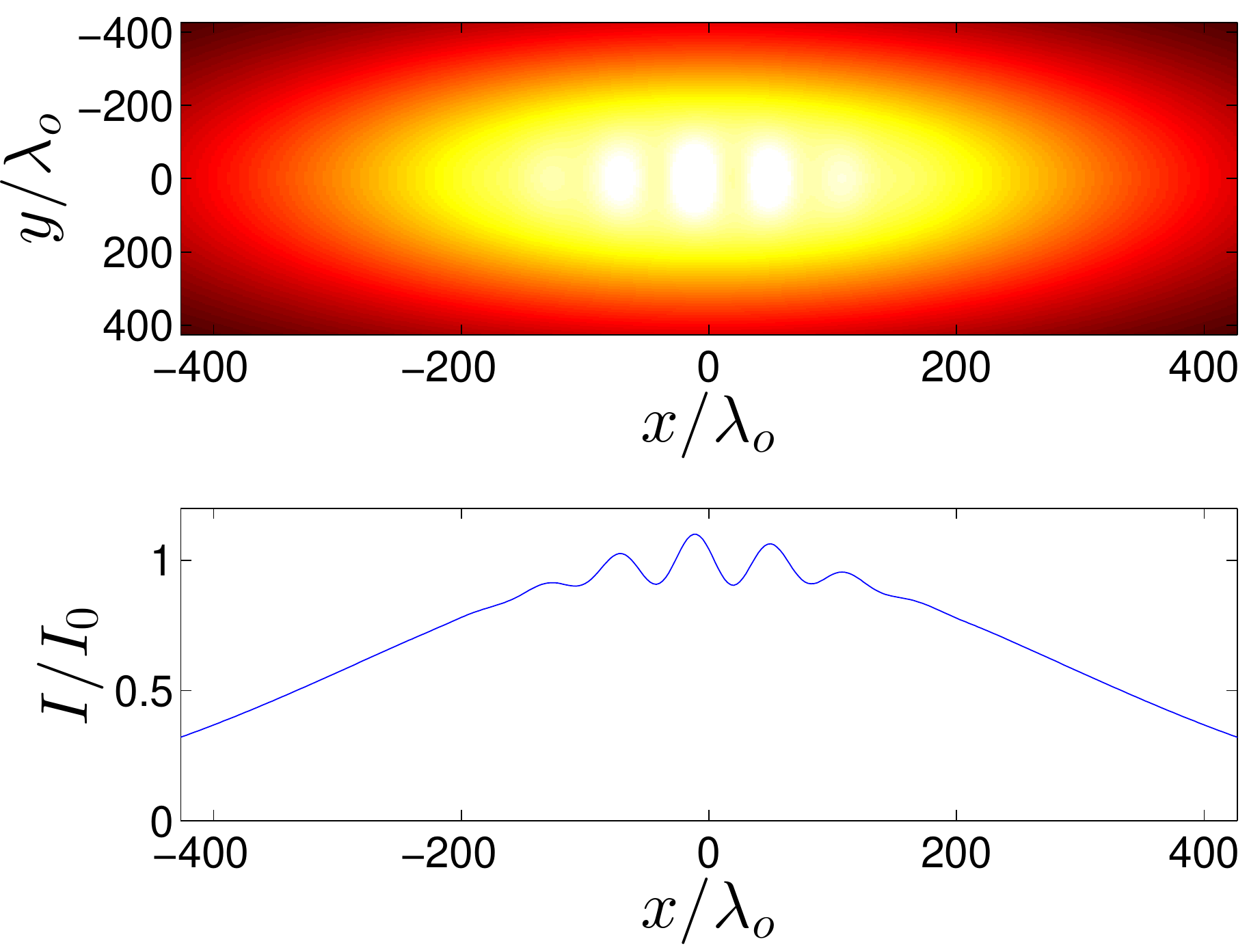}
\includegraphics[width=4cm]{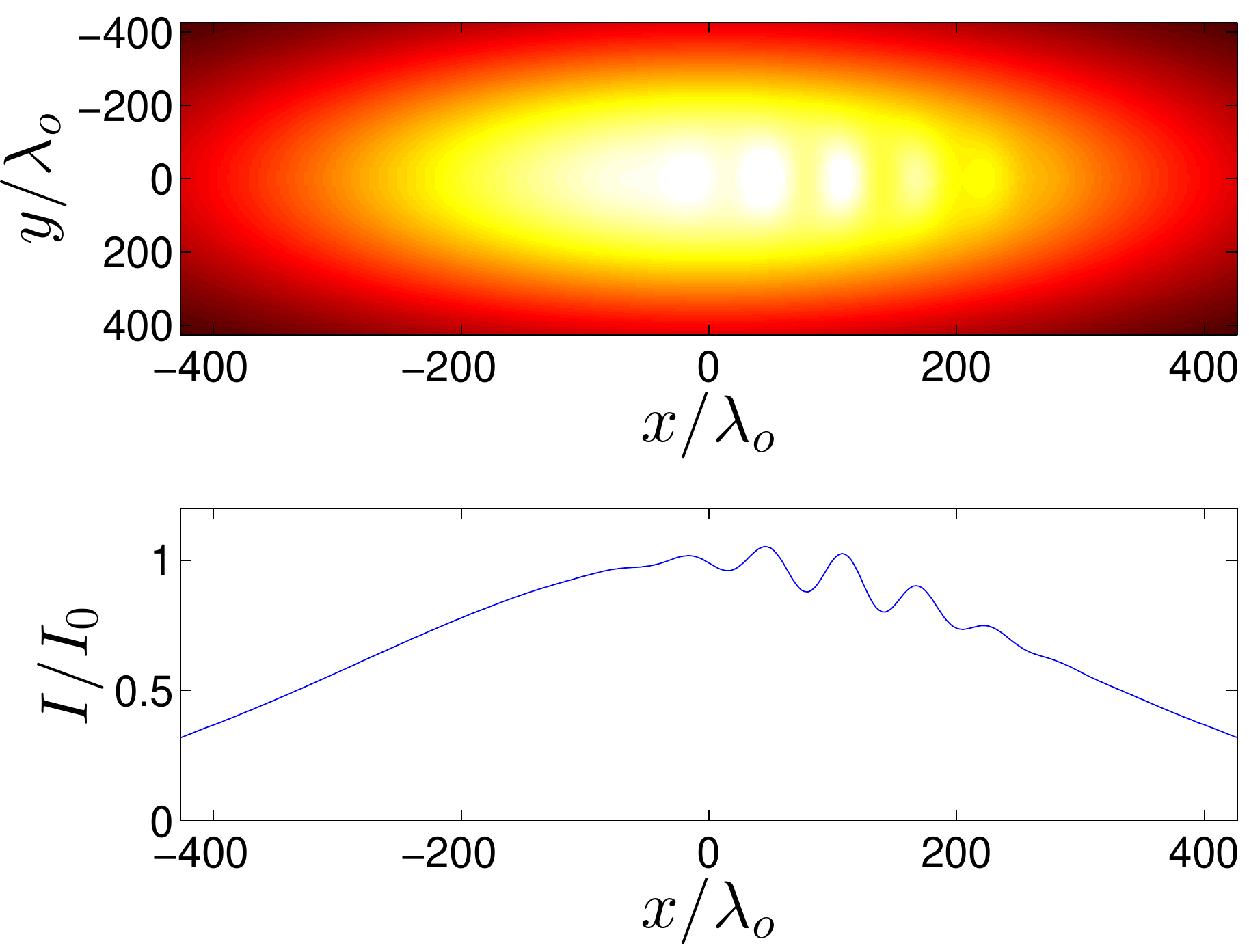}
\includegraphics[width=4cm]{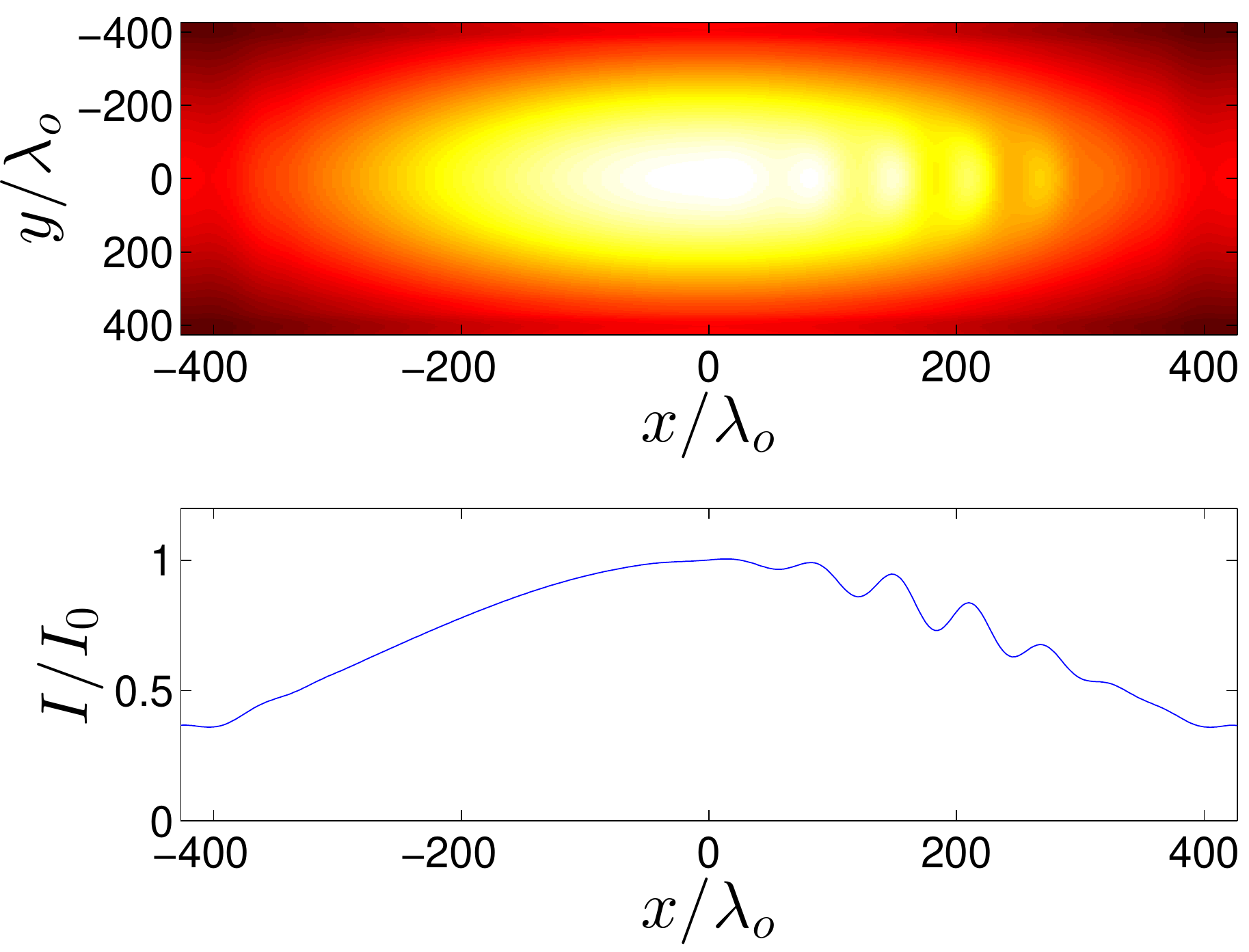}
\caption{Propagation of weak excitations on top of a fluid of light at rest. The incident pump beam is normal to the front crystal interface. Its profile has a overall Gaussian shape with $w_g/\lambda_0=400$. The probe beam has the same frequency and a smaller Gaussian shape with  $w_{\rm pr}/\lambda_0=100$. It is incident on the front interface at a finite angle, corresponding to an in-plane wavevector $k_\perp^{\rm pr}\lambda_0=0.1$ in the positive $x$ direction. 
The upper row shows a strong pump case where $k_\perp^{\rm pr}\xi_0=0.56$ and the excitation is in the phonon regime; from left to right, the snapshots are for $z/\lambda_0=0$, $3000$ and $7000$. 
The bottom row shows the linear optics limit where $k_\perp^{\rm pr}\xi \gg 1$ and the excitation is a single-particle one; from left to right, the snapshots are for $z/\lambda_0=0$, $7000$ and $12000$. On each row, the color scale is normalized to the incident intensity. For each case, the blue line in the lower panel shows a cut of the normalized intensity along the $y=0$ line. 
\label{fig:phonon}}
 \end{figure}

A simplest experiment to characterize the Bogoliubov modes of a fluid of light is to use a strong and wide monochromatic {\em pump} laser beam of amplitude $E_0$ to generate the background fluid of light, and then to use a second, weaker {\em probe} beam at the same frequency $\omega_0$ to create excitations on top of it. This configuration is studied in Fig.\ref{fig:phonon}: the pump spot is taken to have a Gaussian shape and the spatially localized perturbation is created by another Gaussian beam with much smaller waist incident at center of the pump spot with a finite angle $\phi_{\rm pr}$. The in-plane wavevector of the excitations that are created is controlled by the angle $\phi_{\rm pr}$ via the geometric relation $k_\perp^{\rm pr}=(\omega_0/c)\sin \phi_{\rm pr}$.

Because of the peculiar particle-hole mixing of Bogoliubov theory, the excitations that are created at the front interface of the crystal have momenta centered around $\pm \kk_\perp^{\rm pr}$. During propagation, they move with opposite speeds determined by $\vv_{\rm gr}=\nabla_\kk W_{\rm Bog}$ evaluated at $\pm \kk_\perp^{\rm pr}$. Depending on the value of the $k_\perp^{\rm inc} \xi$ parameter, the generated excitation wickll have a phonon (top row in the figure) or single-particle (bottom row) character. In the former case, both components are visible at long times as two separated wavepackets, symmetrically located with respect to $x=0$. In the latter case, the amplitude of the backward component at $\qq=-\kk_\perp^{\rm pr}$ is much smaller than the forward propagating one, and is invisible on the figure.

\section{Suppressed scattering from a localized defect}
\label{sec:defect}

One of the most important consequences of superfluidity is the suppressed drag force felt by a moving impurity crossing the superfluid at slow speeds. In this section, we discuss how an optical analog of superfluidity can be observed in the present context of fluids of light in a propagating geometry. Inspired from previous work on superfluid light in planar cavities~\cite{ICCCRMP,Amo2009,Amo11,Nardin,Sanvitto,ICCCPRL2004} and in atomic BECs~\cite{ICatoms}, we consider a light beam that is moving at a finite speed in the transverse direction and hits a cylindrical defect located around $\rr_\perp=0$. The flow speed in the transverse plane is controlled by the incidence angle $\phi$ of the incident beam used to create the fluid: the transverse wavevector has magnitude $k_\perp^{\rm inc}=(\omega_0/c)\sin \phi$ and corresponds to a transverse flow speed to 
\begin{equation}
v=\epsilon^{-1/2}\,\sin \phi.
\eqname{v}
\end{equation}
The speed of sound $c^0_s$ is controlled by the incident intensity value $|E_0|^2$ via \eqcite{cs}. The defect is described as a Gaussian-shaped modulation of the linear dielectric constant of the form
\begin{equation}
 \delta \epsilon(\rr_\perp,z)=\delta \epsilon_{\rm max} \, \exp (-r_\perp^2/2\sigma^2)
\end{equation}
centered at $\rr_\perp=0$ and of spatial size $\sigma$. 
This may model either a cylindrical region with different chemical or physical properties (e.g. a hole in the glass~\cite{jena}), or the refractive index perturbation generated by an additional laser beam in, e.g., a photorefractive material~\cite{Wan2}.

\begin{figure}[htbp]
 \includegraphics[width=4cm]{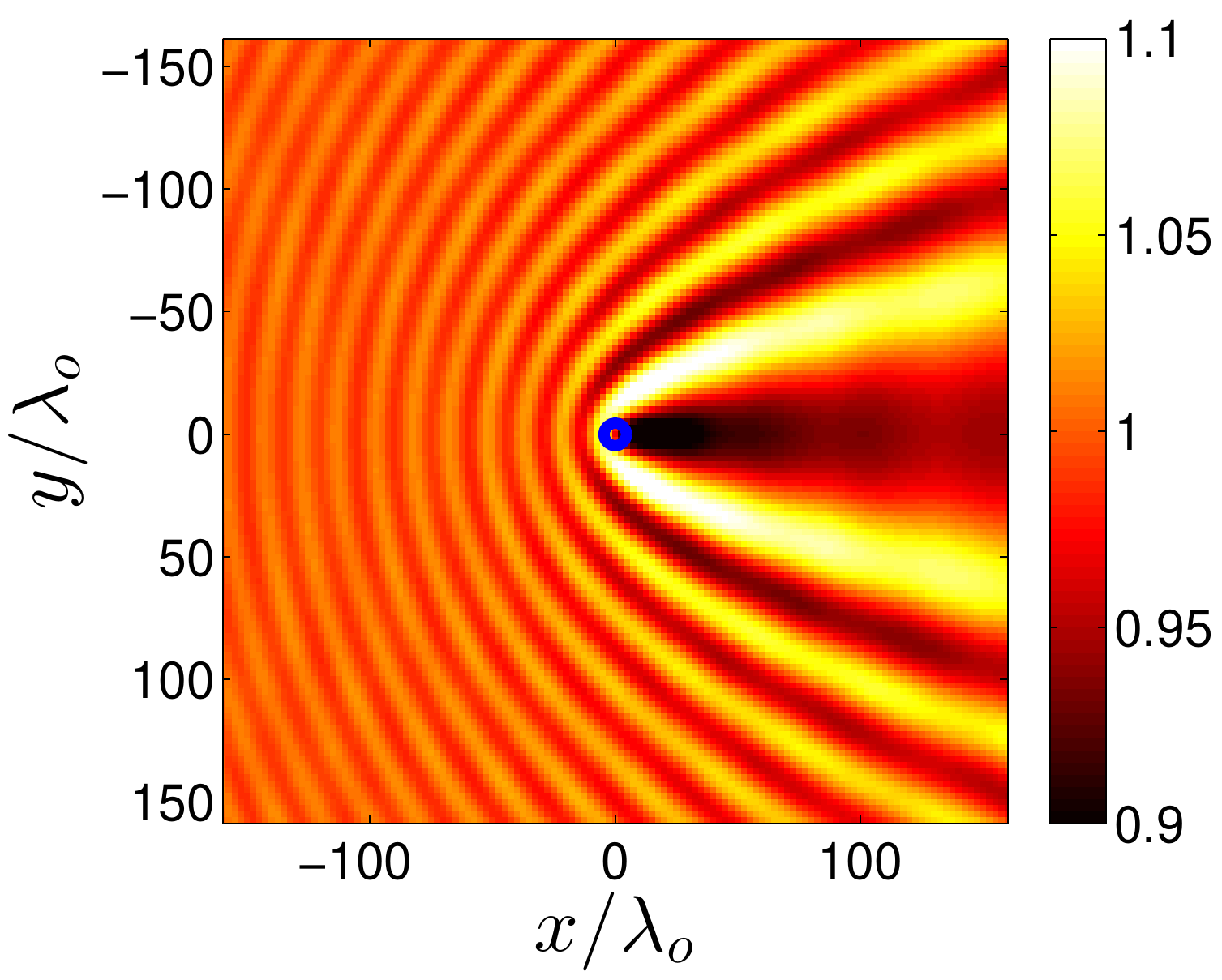}
 \includegraphics[width=4cm]{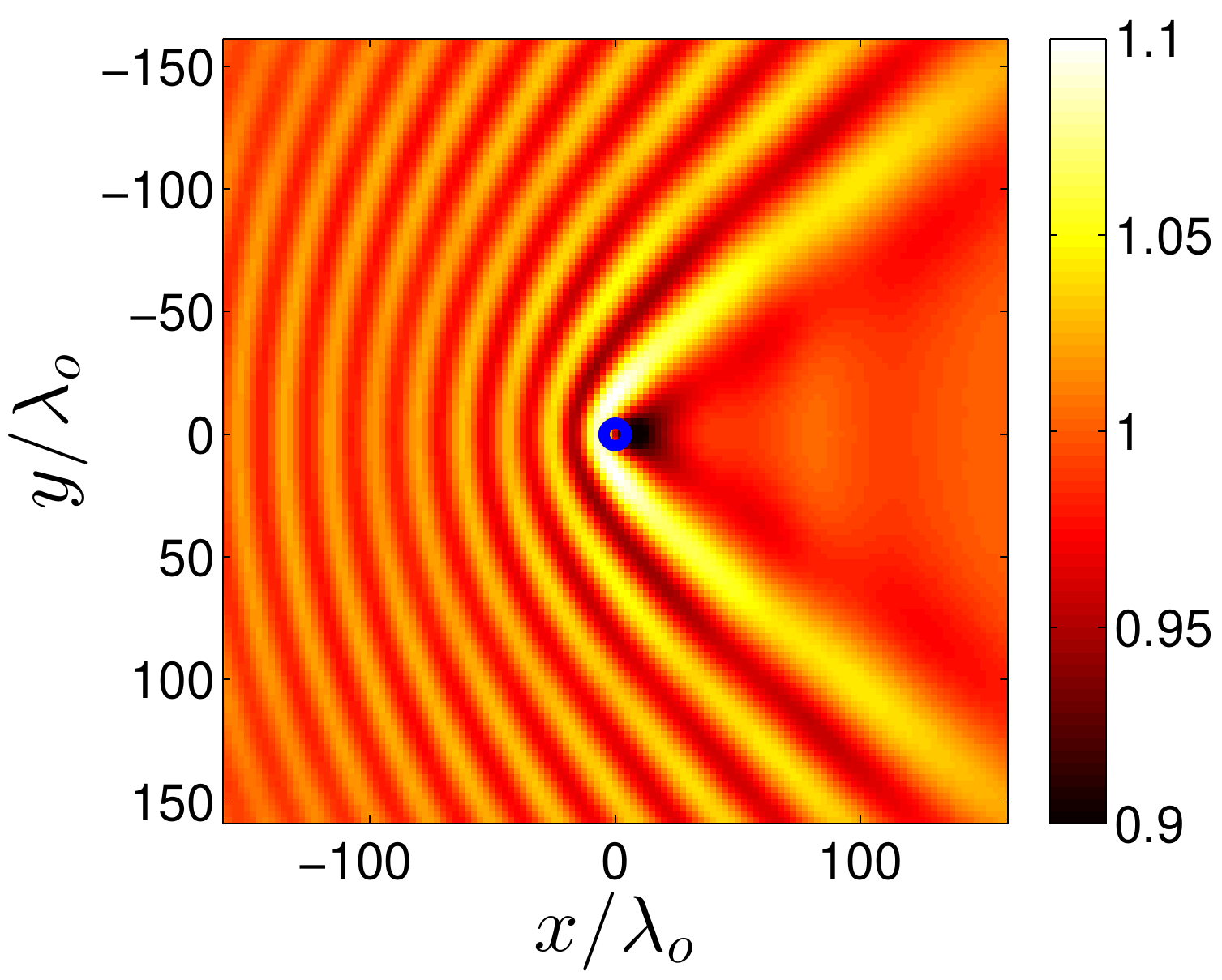} 
\includegraphics[width=4cm]{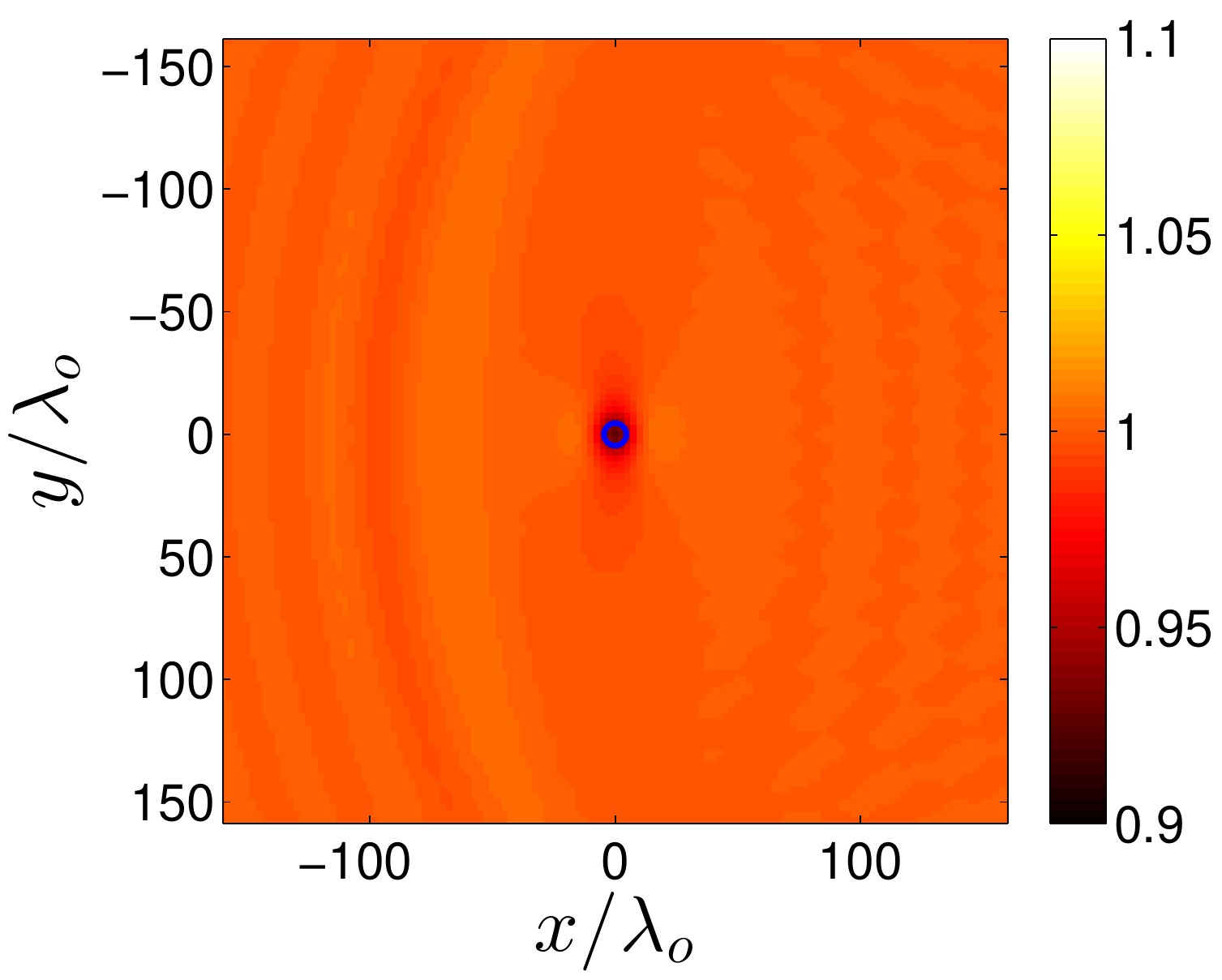}
\caption{Long-distance asymptotic transverse profiles of the laser beam intensity hitting a cylindrical defect located at $\rr_\perp=0$. 
The flow velocity is the same $v=0.034$ along the positive $x$ direction (the right-ward direction in the figure), while the light intensity is different in the three panels, increasing from left to right. The corresponding Mach numbers are $v/c^0_s=\infty$ (linear optics limit, left panel), $v/c^0_s=1.84$ (supersonic flow regime, central panel), $v/c^0_s=0.86$ (superfluid regime, right panel).
The defect parameters are $\sigma/\lambda_0=5$ and $\delta\epsilon_{\rm max}/\epsilon=-1.6\cdot10^{-4}$. 
The color map is normalized to the incident intensity. 
\label{fig:superfl_big}}
 \end{figure}

We begin our discussion from the weak defect regime where the small perturbation induced in the fluid can be described within a linearized Bogoliubov theory. Fig.\ref{fig:superfl_big} illustrates the main regimes in the geometrically simplest case where the incident beam has a very wide profile. In this case, a $z\to \infty$ long-distance limit can be taken where the beam has already propagated for a very long distance and all transients have disappeared. Far from the defect, the light intensity profile recovers the incident intensity value $|E_0|^2$.

The right panel shows the superfluid regime $v<c^0_s$: the fluid of light moves at a sub-sonic speed and is able to flow around the defect without any friction. This is visible in the absence of any intensity modulation far from the defect. The central panel shows a supersonic flow regime where $v>c^0_s$ and superflow is broken: a Mach-Cerenkov cone appears downstream of the defect, with an aperture $\sin\theta=c_s^0/v$; in addition, parabolic-like precursors appear upstream of the defect. The left panel shows the extremely supersonic regime with $v\gg c^0_s$ that occurs in the linear optics regime at very low incident intensities: in this case, the linear interference of the incident and scattered light is responsible for parabolic shape of the fringes~\cite{duck}. 

Given the simple relation \eqcite{v} between the incidence angle and the flow speed, the sub-sonic $v<c_s^0$ condition for superfluid flow translates in the present propagating geometry into a condition on the (small) incidence angle 
\begin{equation}
 \phi<\sqrt{\epsilon}\,c_s^0=\left[-\frac{\chi^{(3)}\,|E_0|^2}{2}\right]^{1/2}.
\end{equation}

\begin{figure}[htbp]
 \includegraphics[width=4cm]{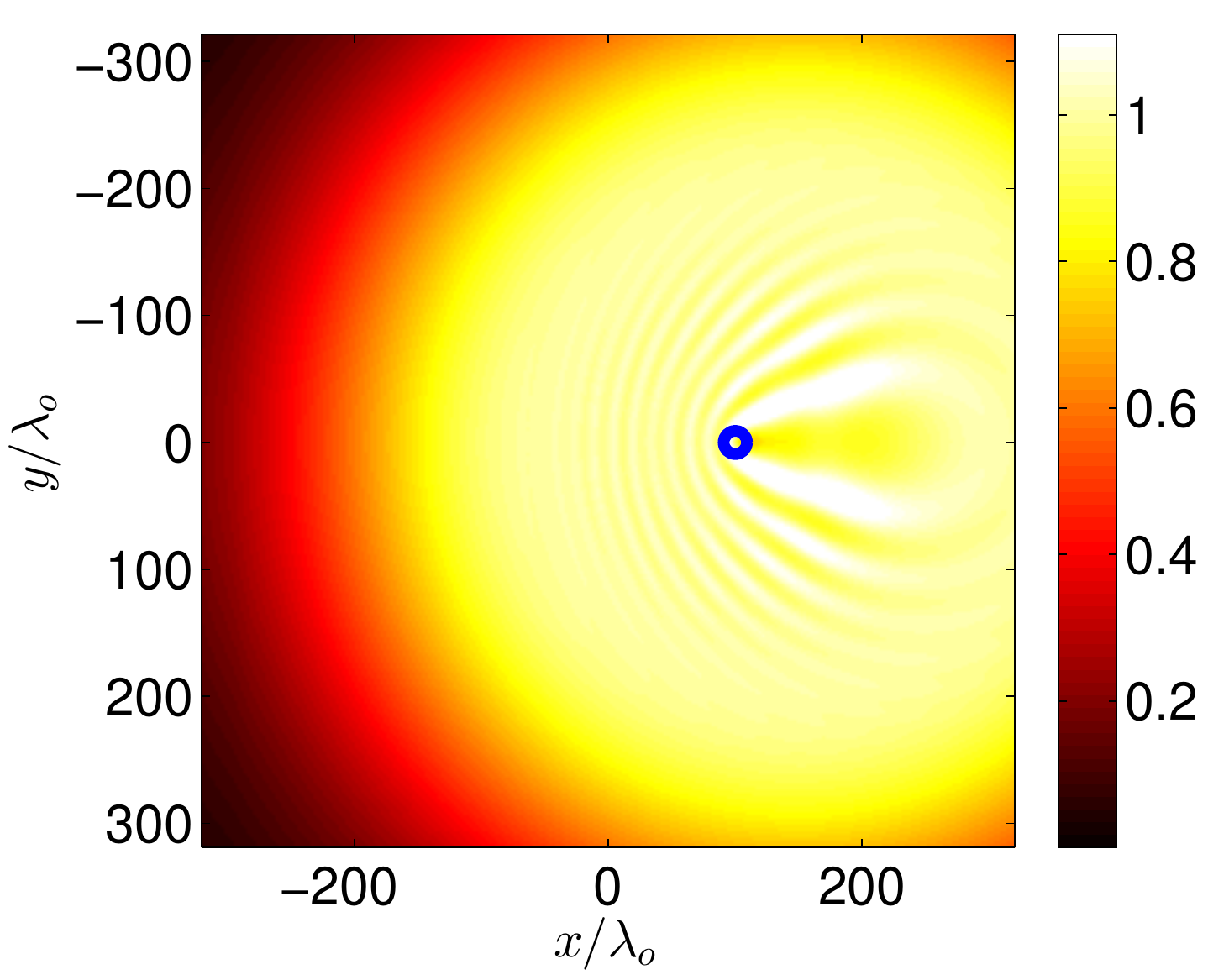}
 \includegraphics[width=4cm]{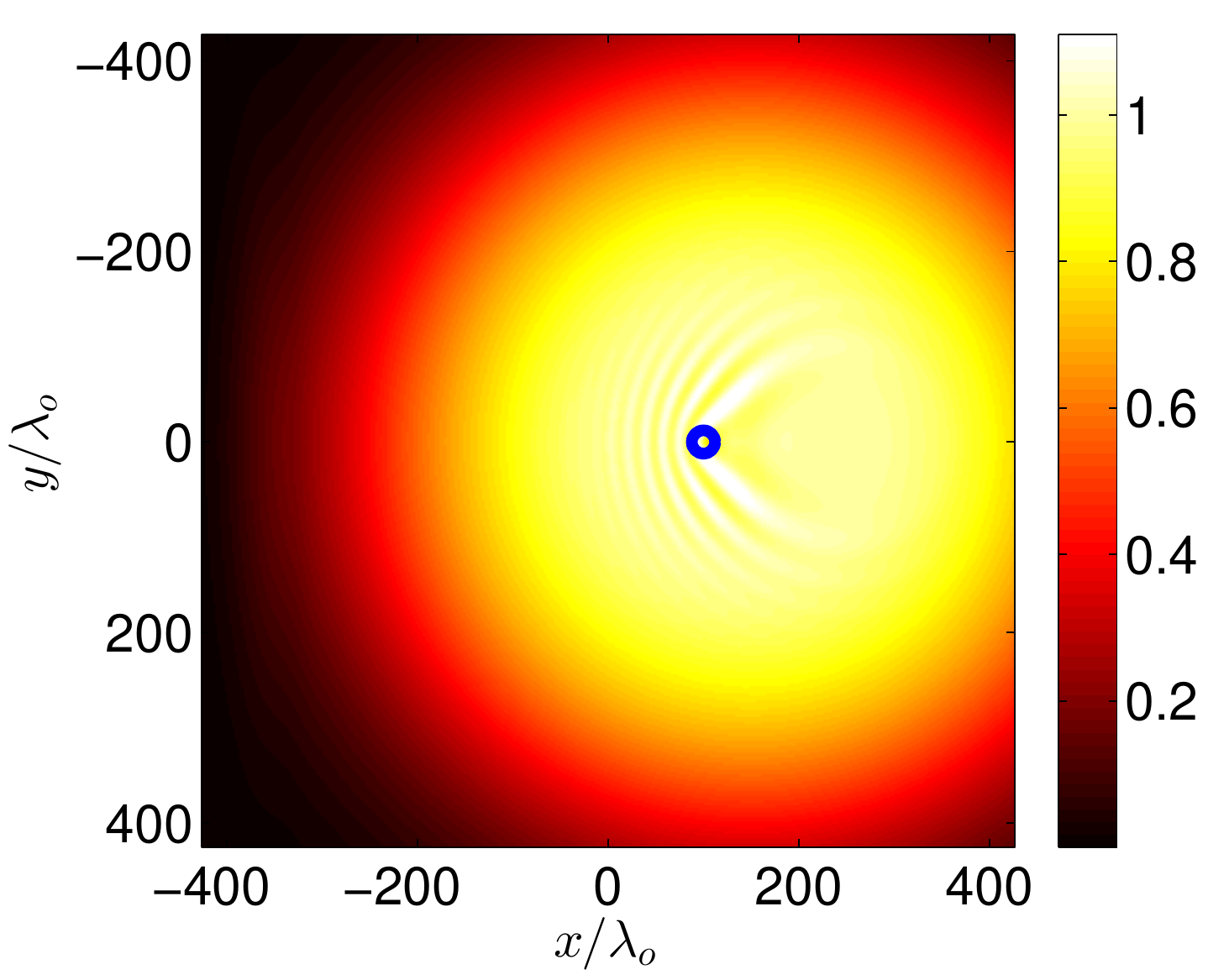}
  \includegraphics[width=4cm]{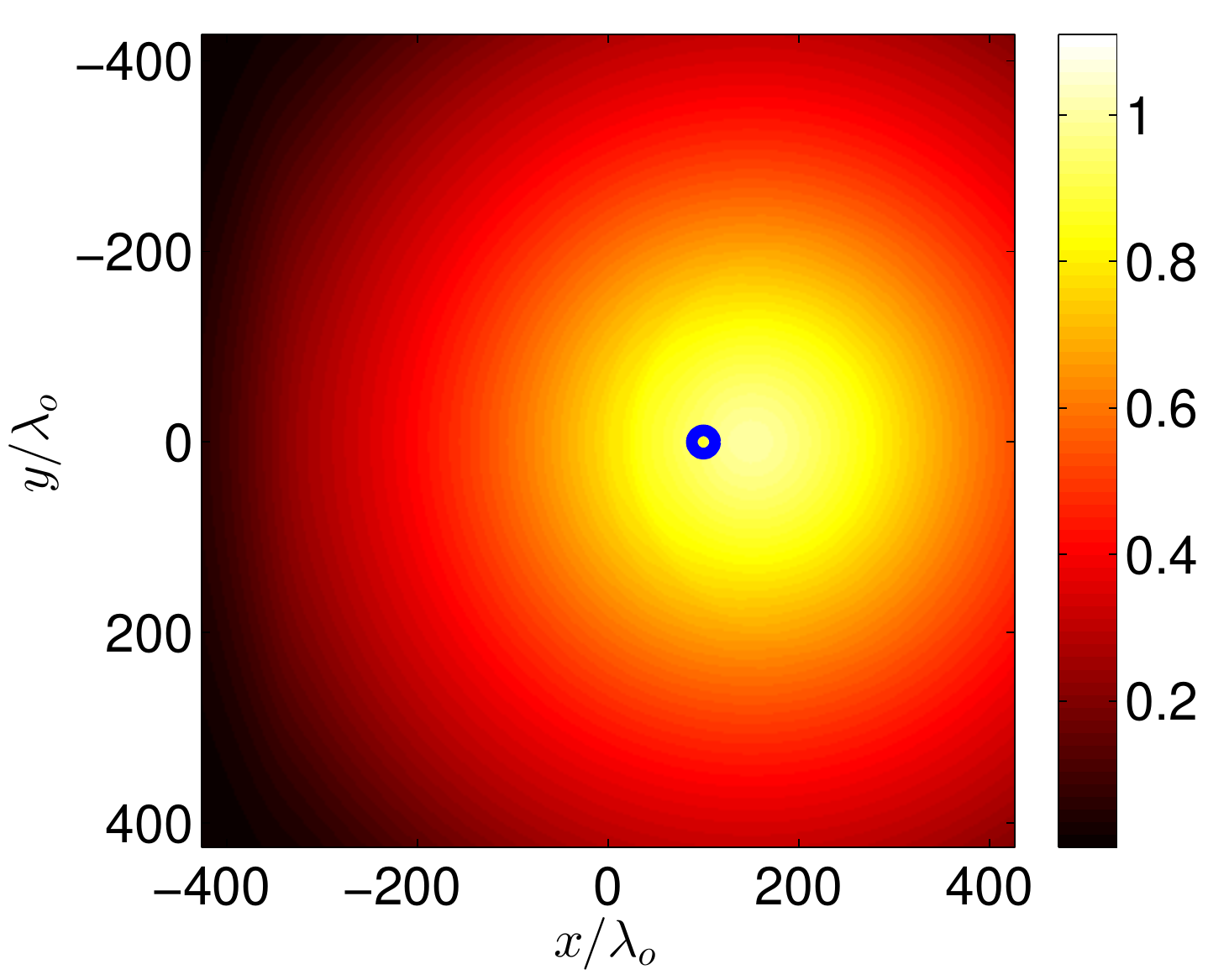}
\caption{
Transverse profiles of the laser beam intensity after a propagation distance $L/\lambda_0=4500$ in the nonlinear crystal. The incident beam has the flat-top intensity profile shown in the left panel of Fig.\ref{fig:superfl_tdep} and a transverse carrier wavevector giving a flow velocity $v=0.034$ along the positive $x$ direction. As in Fig.\ref{fig:superfl_big}, the peak light intensity increases from left to right: the left panel shows the linear optics regime with $v/c^0_s\simeq \infty$, the central panel shows the supersonic flow regime $v/c^0_s=1.88$, the right panel shows the superfluid regime $v/c^0_s=0.77$. Same defect parameters as in Fig.\ref{fig:superfl_big}.
\label{fig:superfl}
}
 \end{figure}

\begin{figure}
 \includegraphics[width=4cm]{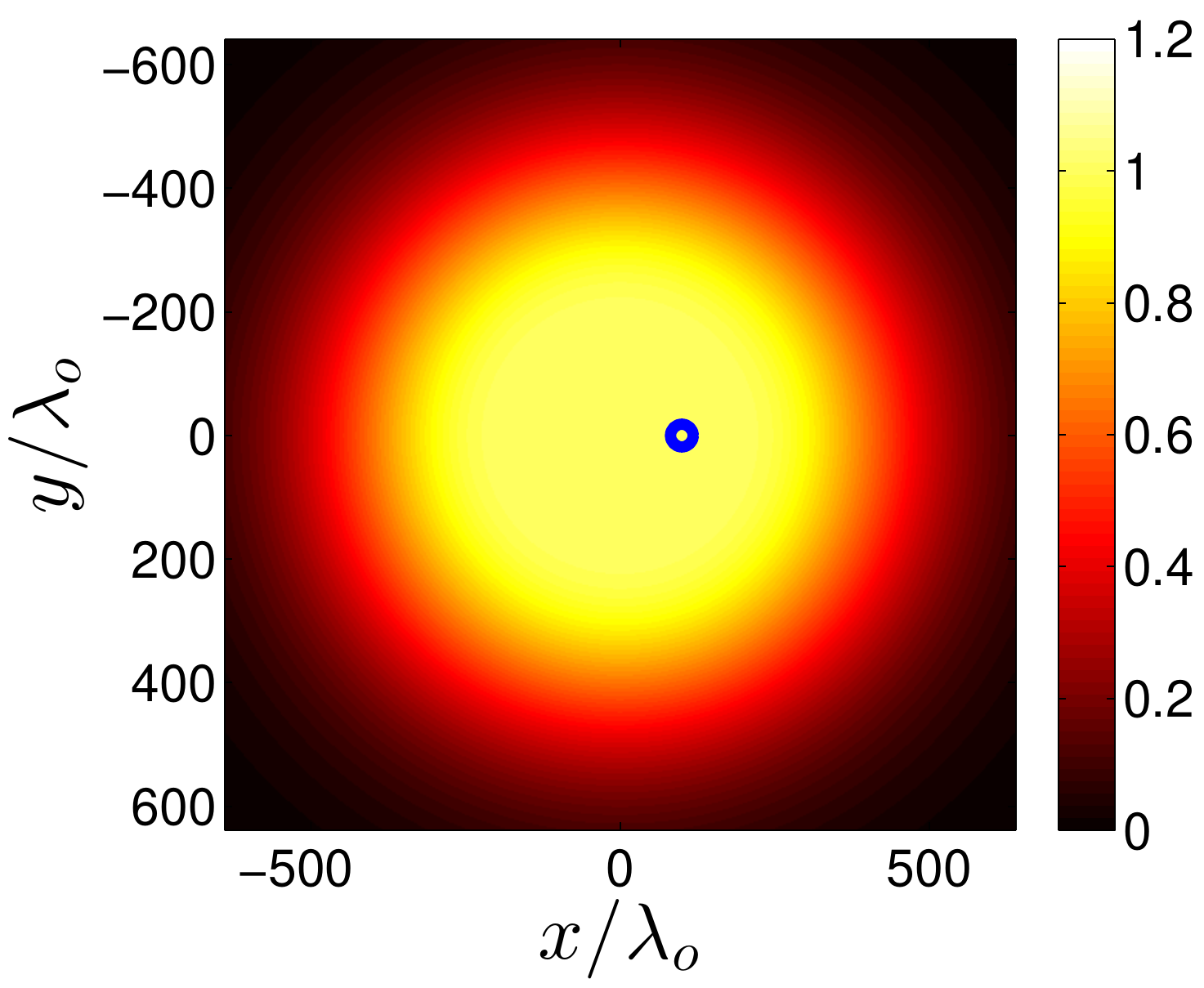}
 \includegraphics[width=4cm]{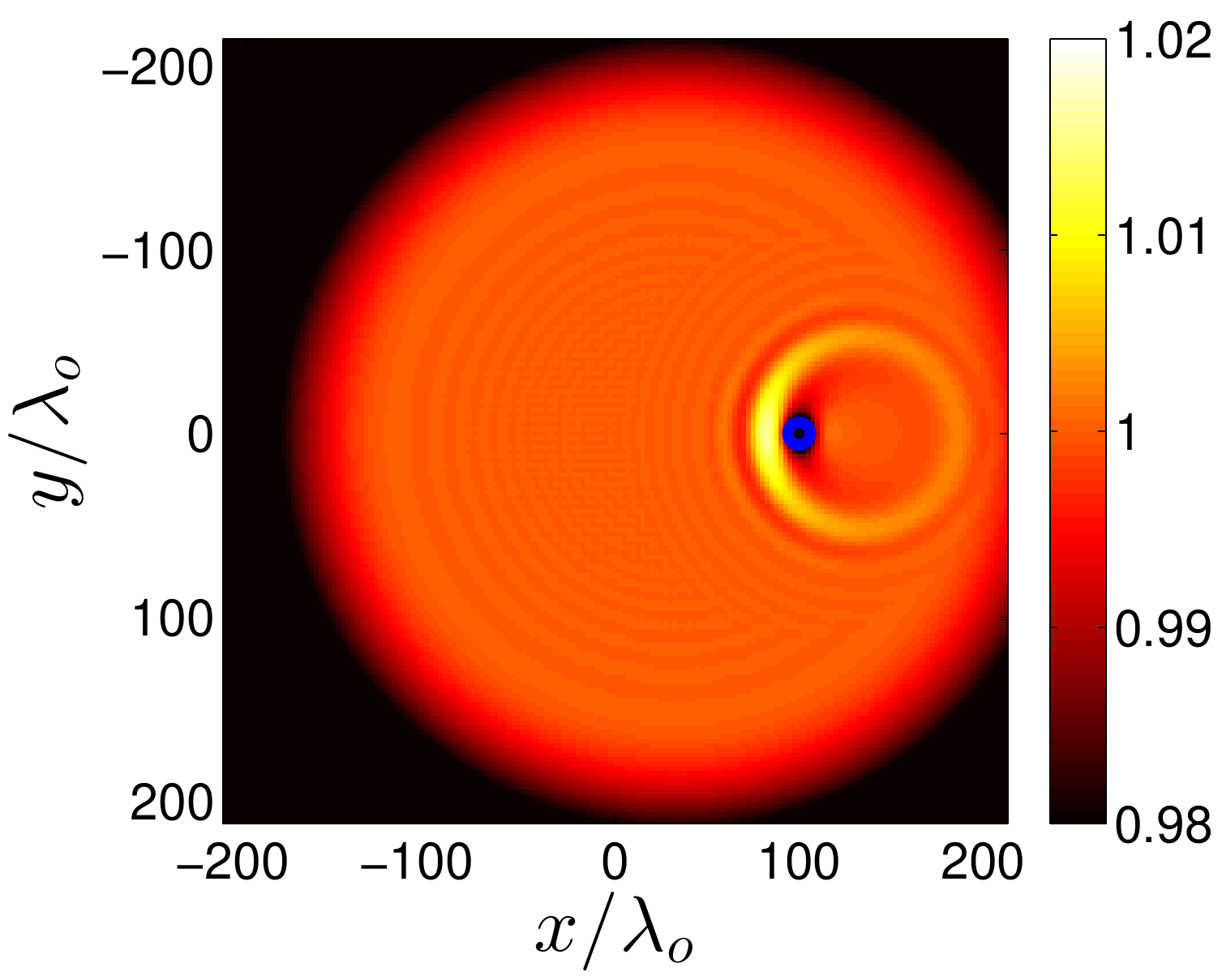}
 \includegraphics[width=4cm]{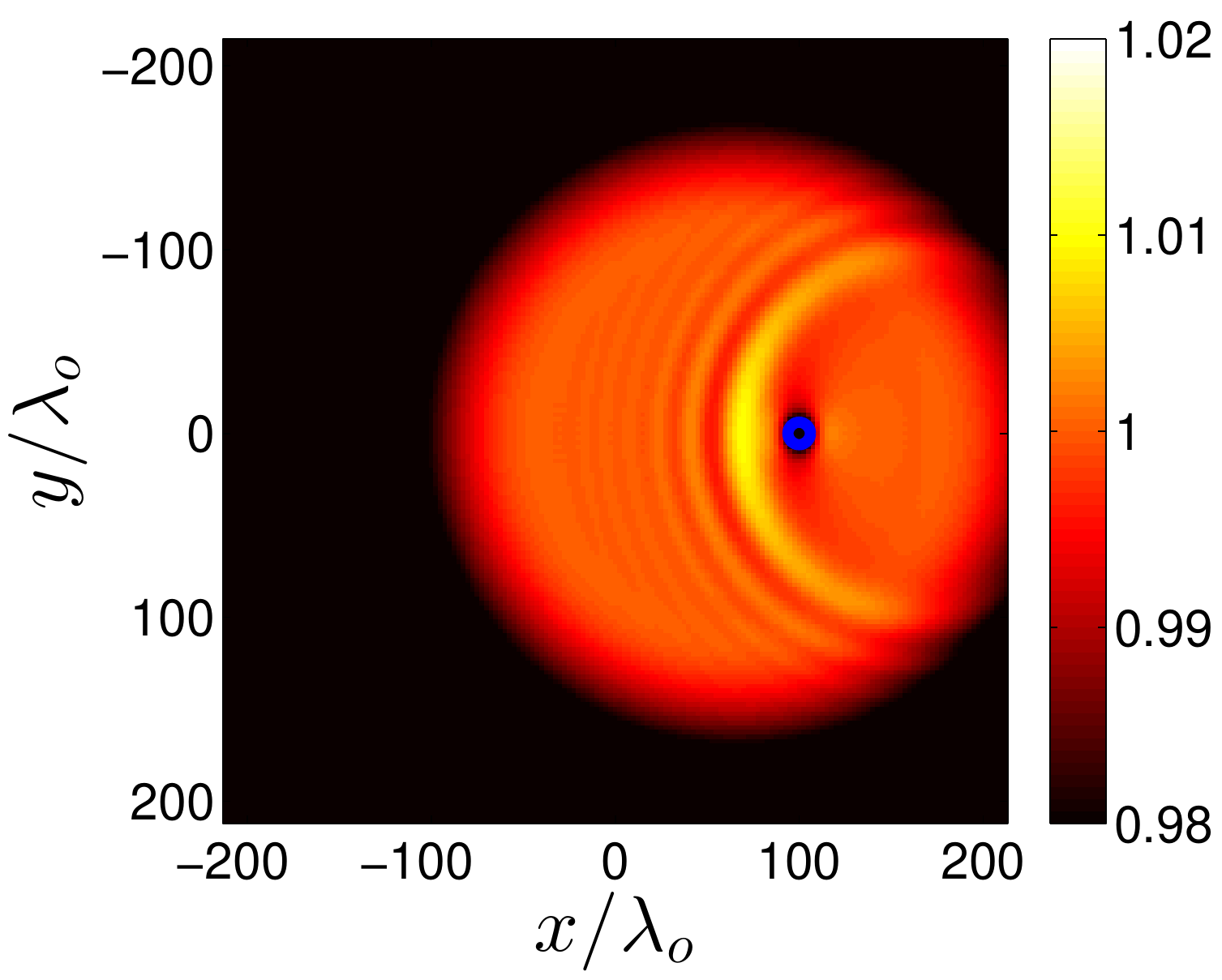}
\caption{The left panel shows the transverse intensity profile of the flat-top incident laser beam: the inner flat region has a radius $w_f/\lambda_0=200$ and the Gaussian wings extend for $w_g/\lambda_0=300$. The central and right panels show the transverse intensity profiles after propagating for $L/\lambda_0=1000$ (center) and $2000$ (right). Same configuration as in the right panel of Fig.\ref{fig:superfl}, but note the different color scale. 
\label{fig:superfl_tdep}}
 \end{figure}

Figs.\ref{fig:superfl} illustrates how this physics is modified for realistic values of the crystal size and incident pump waist. In particular, we consider the flat-top incident beam of peak amplitude $E_0$ shown in Fig.\ref{fig:superfl_tdep}(a). As before, the super- or sub-sonic nature of the flow is defined according to the peak sound speed $c_s^0$ determined by inserting the peak intensity $|E_0|^2$ into \eqcite{cs}. During propagation, the beam spot globally moves with speed $v$ in the rightward direction and suffers some spatial expansion under the effect of the repulsive interactions. 

In the supersonic case shown in the left and central panels of Fig.\ref{fig:superfl}, the main visible difference with respect to the corresponding panels of Fig.\ref{fig:superfl_big} is the finite spatial extension of the modulation pattern originating from the defect: the modulation starts forming as soon as light interacts with the defect and its spatial size keeps growing during propagation. At a given position, the modulation tends to a constant shape corresponding to the asymptotic pattern shown in Fig.\ref{fig:superfl_big}. 

In superfluid regime shown in the right panel of Fig.\ref{fig:superfl} and, in more detail, in the central and right panels of Fig.\ref{fig:superfl_tdep} there is also a visible transient that propagates from the defect as a spherical wave. This wave is generated when the initially unperturbed Gaussian spot first hits the localized defect. While the whole spherical pattern drifts laterally as a consequence of the overall flow speed $v$, the radius of its inner rim grows at the speed of sound $c_s$ and shorter wavelength precursors expand at a faster rate as a consequence of the super-luminal nature of the Bogoliubov dispersion. Once the spherical wave has moved away from the defect, the intensity pattern tends to the asymptotic pattern shown in the right panel of Fig.\ref{fig:superfl_big} which only shows a localized density dip at the defect location. 

\begin{figure}
\includegraphics[width=4cm]{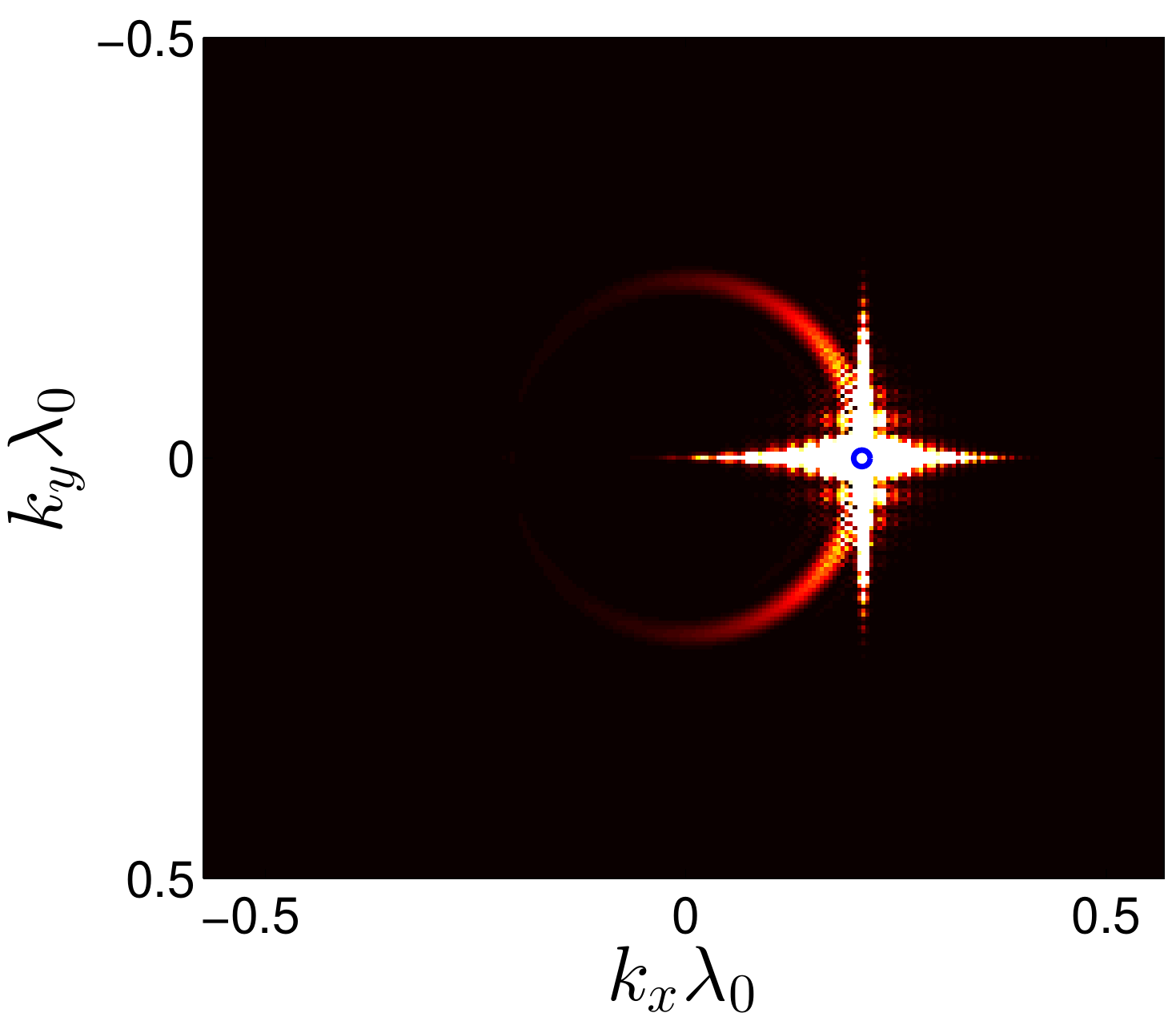}
 \includegraphics[width=4cm]{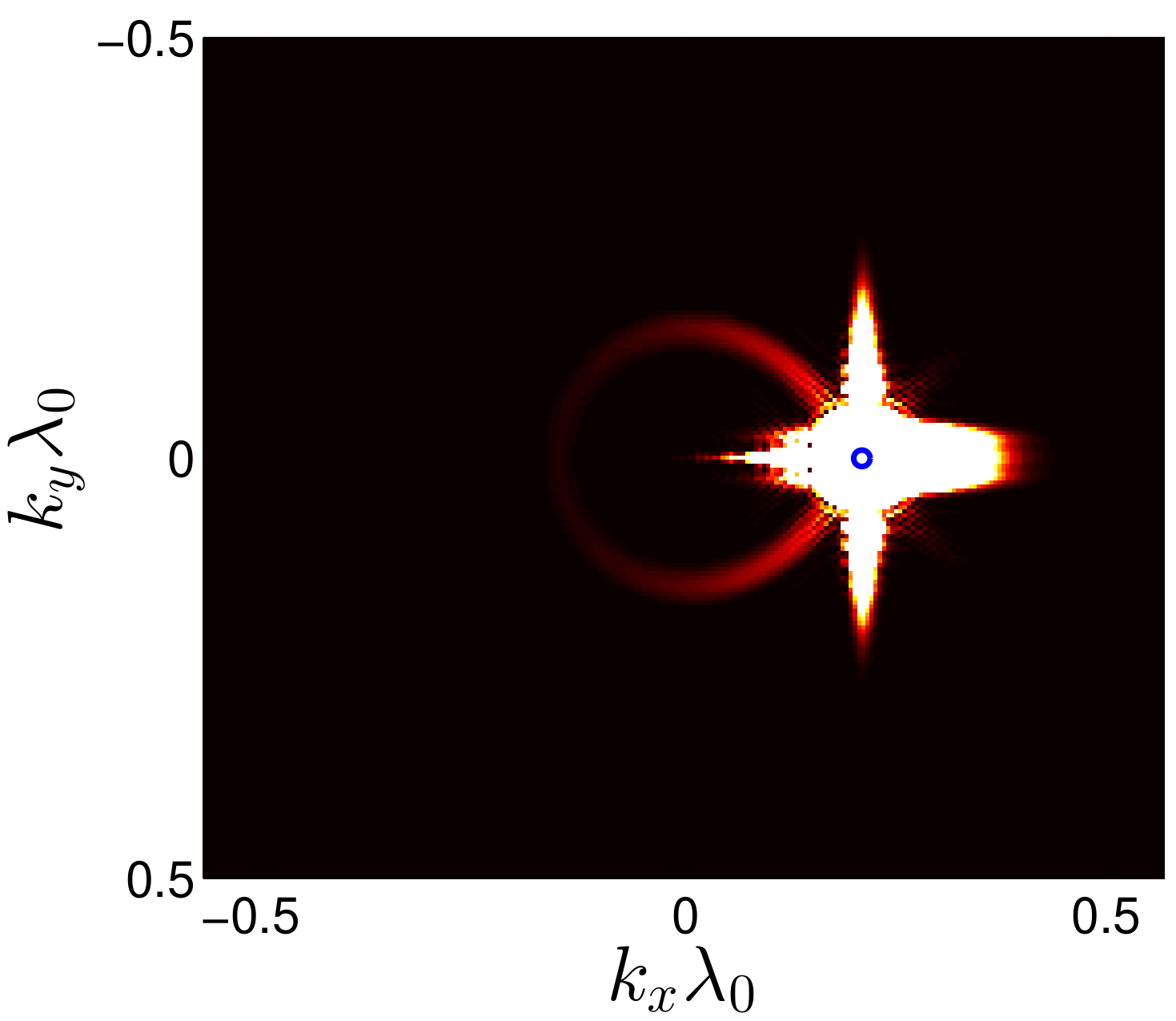} 
 \includegraphics[width=4cm]{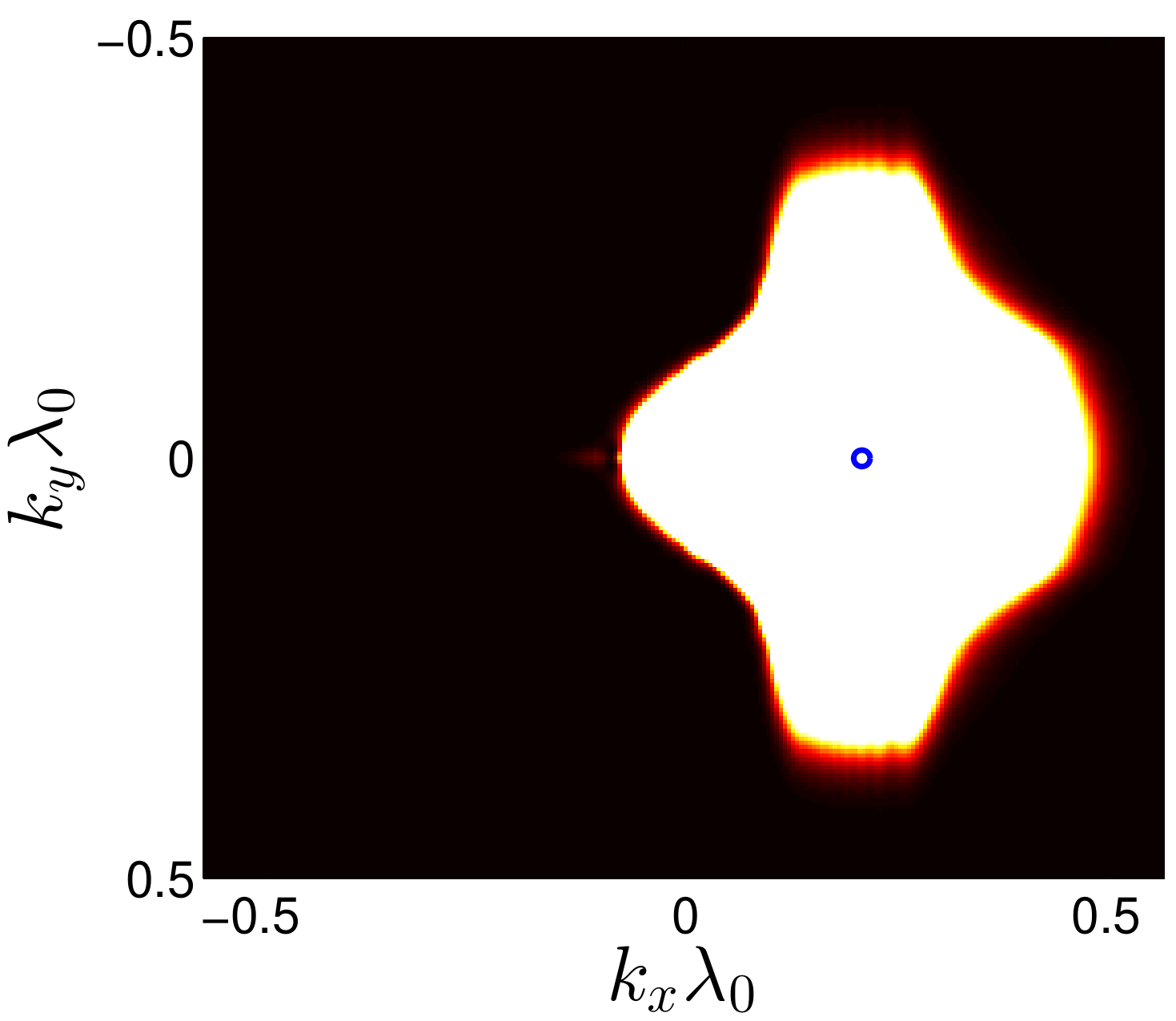}
\caption{Far field emission patterns for the same configurations as in Fig.\ref{fig:superfl}. The blue circle indicates the incident wavevector $\kk_\perp^{\rm inc}$. The wavevector $\kk_\perp$ is related to the emission angle by $\sin \phi_{\rm em}=(k_\perp\lambda_0)\,\sqrt{\epsilon}/2\pi$.
\label{fig:superfl_k}}
 \end{figure}
 
To complete the picture, it is interesting to display also $\kk_\perp$-space profiles of the field amplitude that emerges from the back face of the crystal: these patterns are directly accessible in an experiment as the far field angular pattern of the transmitted light. In the strongly supersonic case $c^0_s\ll v$ shown in the left panel, scattering on the defect is responsible for a ring-shaped feature passing through the incident wavevector $\kk_\perp^{\rm inc}$. When $c_s$ grows towards $v$, the ring is deformed developing a corner at $\kk_\perp^{\rm inc}$ and a weaker copy of it appears at symmetric position with respect to $k_\perp^{\rm inc}$. This latter feature is another manifestation of the Bogoliubov transformation underlying the definition of the quasi-particle operators. In the $c_s>v$ superfluid regime shown in the right panel, the ring disappears and only a single peak at $\kk_\perp^{\rm inc}$ remains visible: the strong broadening of this peak that is apparent in the figure is due to the overall rapid expansion of the spot under the effect of the repulsive interactions.

\begin{figure}
\includegraphics[width=4cm]{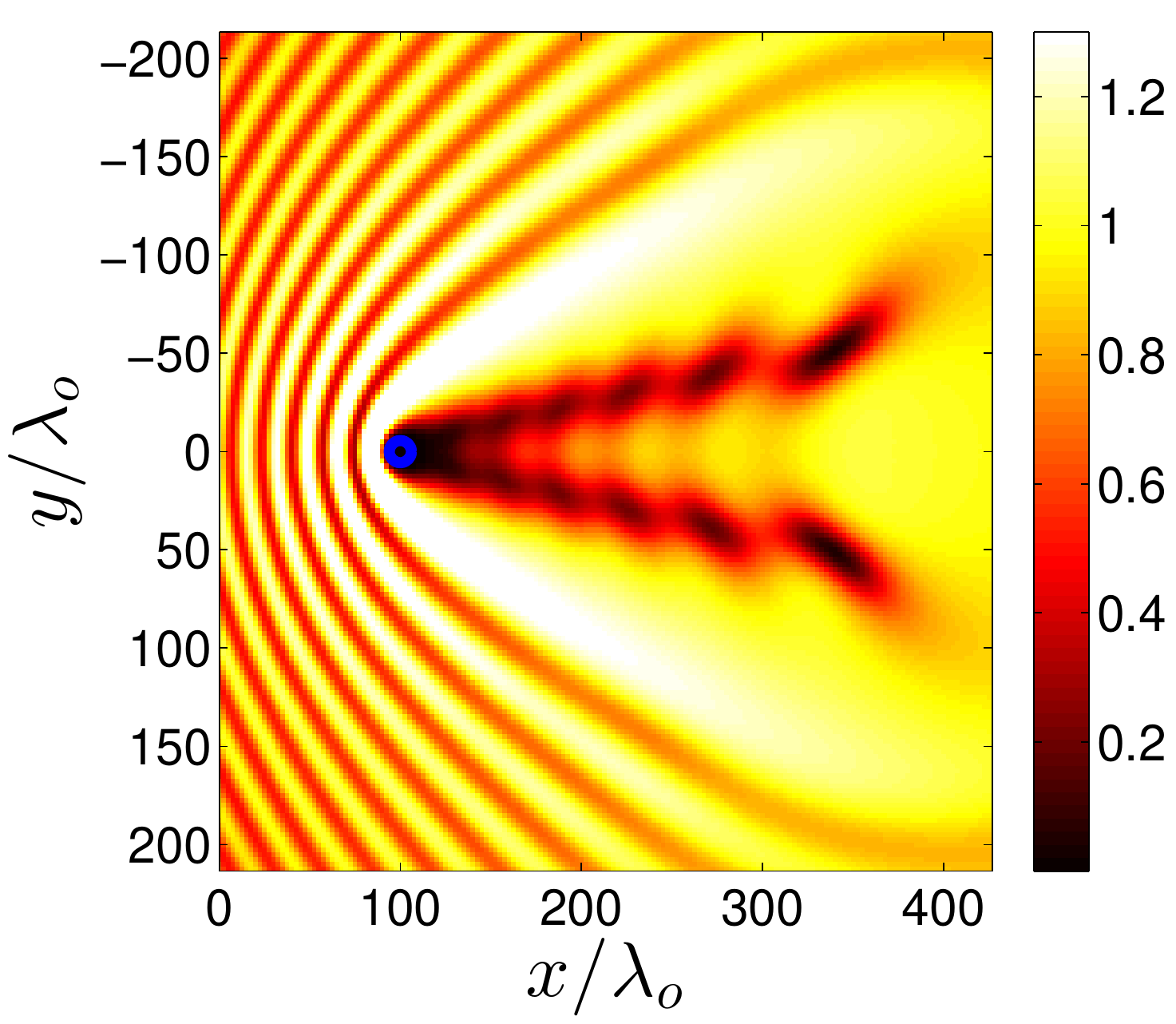}
\includegraphics[width=4cm]{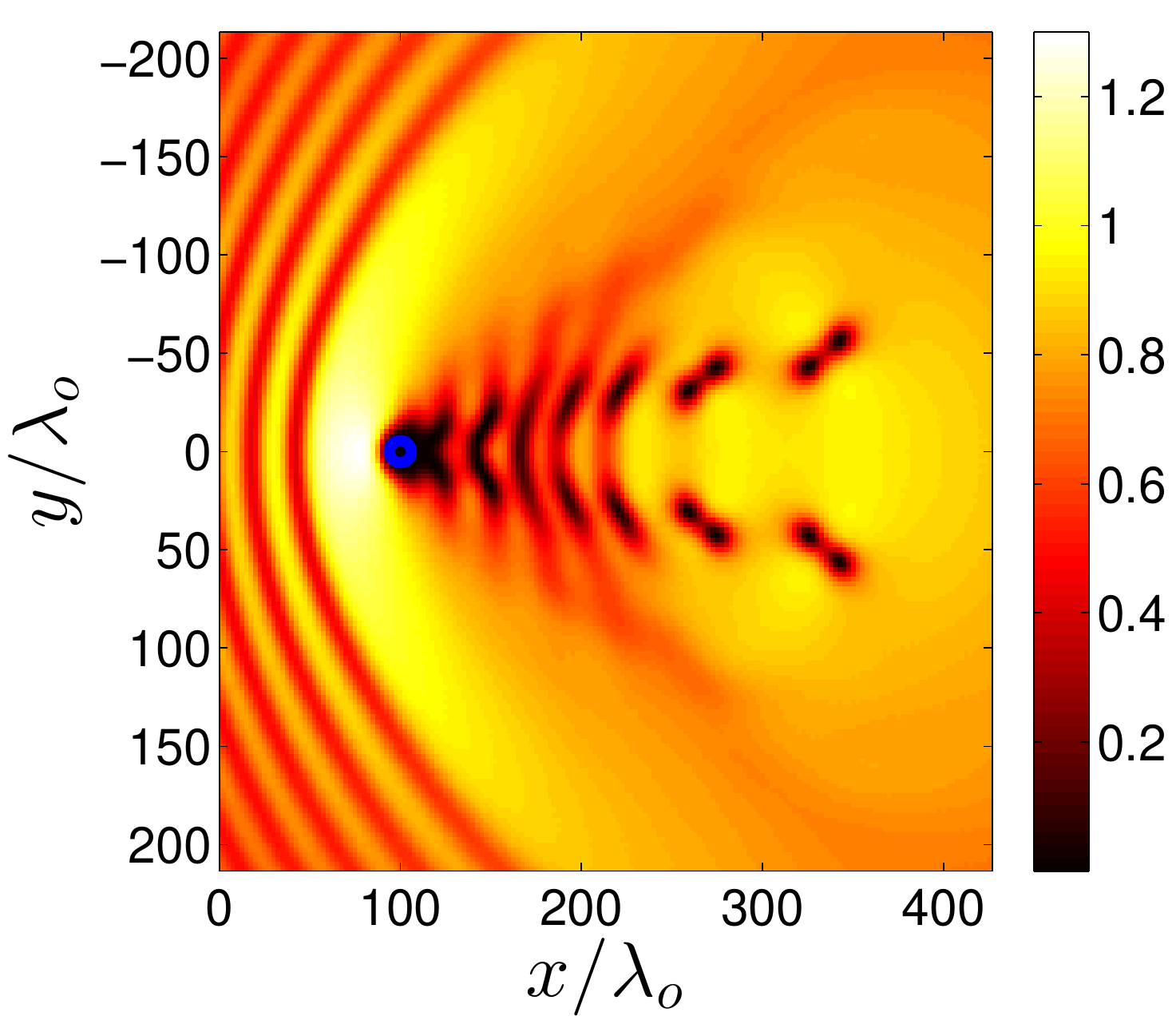}
\includegraphics[width=4cm]{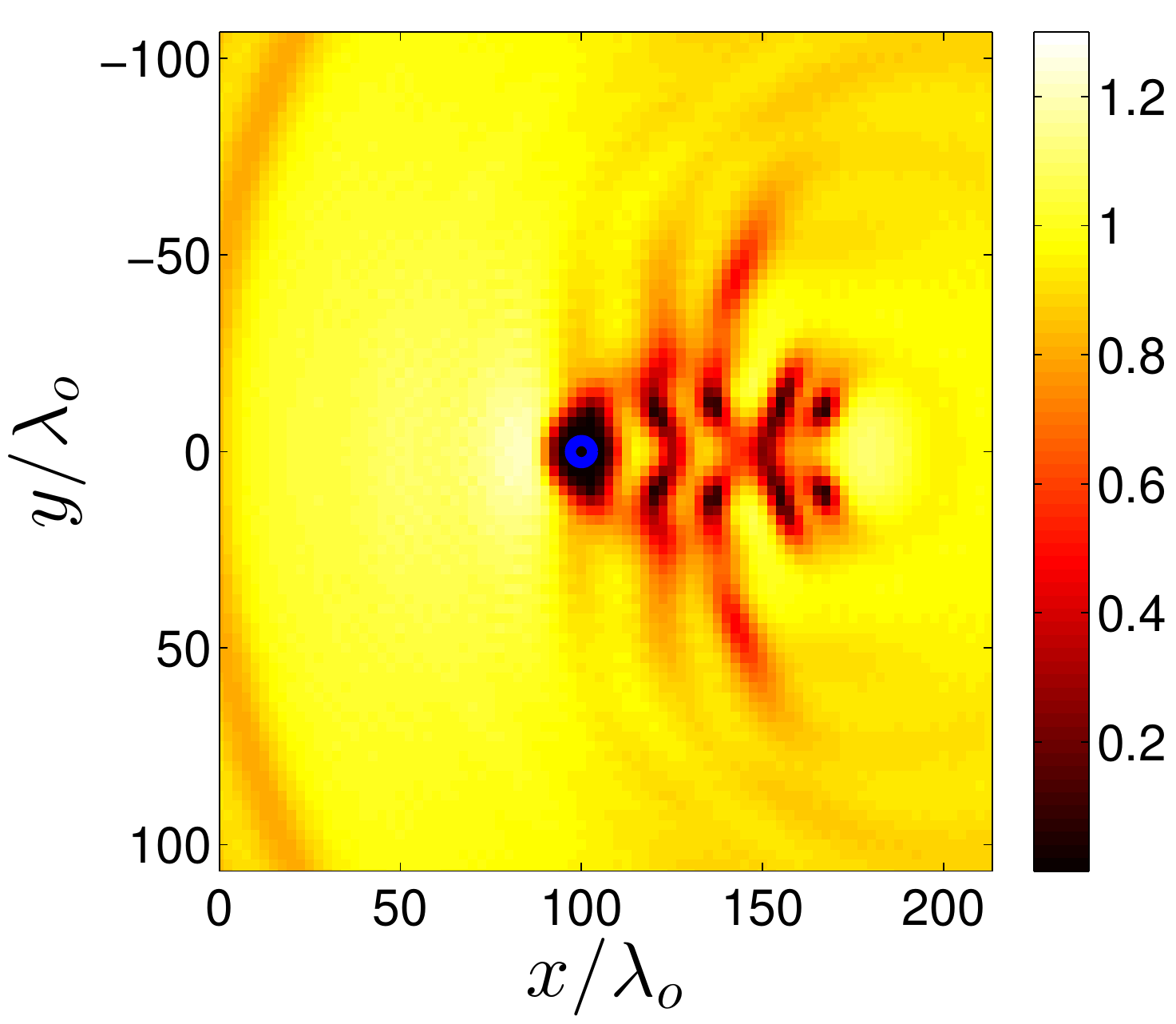}
\caption{Transverse intensity profile of the beam after hitting a strong cylindrical defect with $\sigma/\lambda_0=4$ and $\delta\epsilon_{\rm max}/\epsilon=-0.03$ centered at $\rr_\perp=0$. The flow speed is the same $v=0.034$ in the three panels, while the peak incident intensity grows from left to right, giving $v/c^0_s=2.66$ (left), $v/c^0_s=1.33$ (center) and $v/c^0_s=0.77$ (right). 
Propagation distance $z/\lambda_0=9500$ (left, center) and $4500$ (right). Same flat top incident beam as in Fig.\ref{fig:superfl}.
\label{fig:soliton}}
 \end{figure}
 
As it was originally predicted in the context of superfluid liquid Helium and recently experimentally observed in superfluids of light in planar cavities~\cite{Amo11,Nardin,Sanvitto}, more complex behaviors including the nucleation of solitons and vortices are observed for large and strong defects. First mentions of this physics in the optical context were given in~\cite{opticsvortex}. 
A glimpse of this physics is given in Fig.\ref{fig:soliton} where some most significant examples of the spatial profile of the field after propagation in the nonlinear crystal are shown. For very low speeds, one would recover the superfluid behavior already seen above (not shown). For intermediate speeds on the order of a fraction of $c_s^0$, pairs of vortices are continuously emitted by the defect (right panel). For large speeds $v>c_s^0$, the vortices tend to merge and form a pair of oblique dark solitons (left panel).

\section{Trans-sonic flows}
\label{sec:BH}

In this last Section, I wish to briefly present a novel research axe where the remarkable properties of superfluid light in propagating geometries could be exploited to experimentally investigate aspects of quantum field theory on a curved space-time in a novel context.

Following the early theoretical work \cite{Fouxon}, the pioneering experiment in \cite{BarAd} has demonstrated a trans-sonic flow configuration in a fluid of light: using a spatial constriction, an interface can be created in the flowing light which separates an upstream region of sub-sonic flow with $c_s>v$ from a downstream one of super-sonic flow $v>c_s$. As discussed at length in the literature on the so-called {\em analog models}~\cite{Barcelo}, the trans-sonic interface between the two regions shares close similarities with a black-hole horizon in gravitational physics.

\begin{figure}
\begin{center}
\includegraphics[width=3cm]{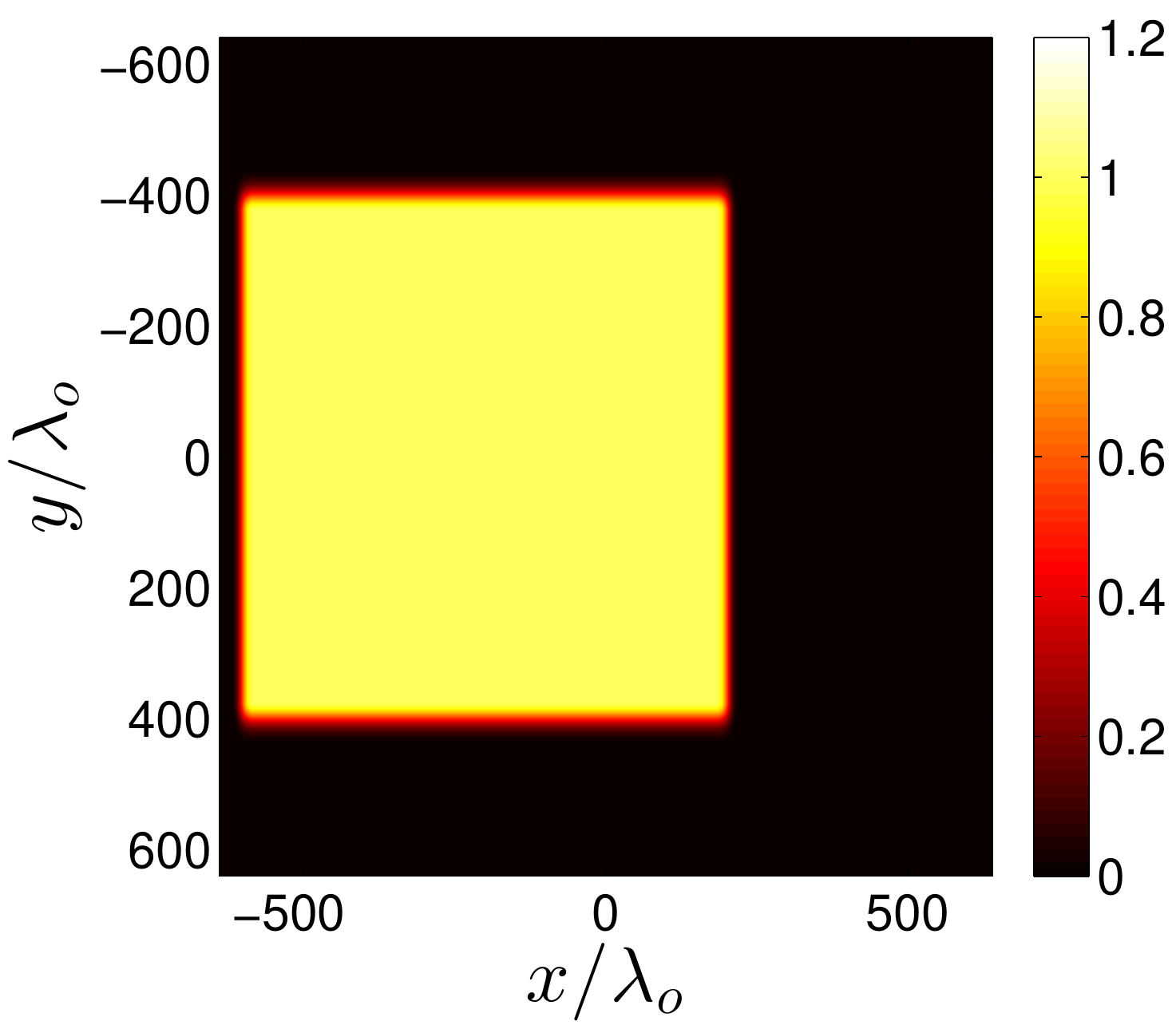}
\includegraphics[width=3cm]{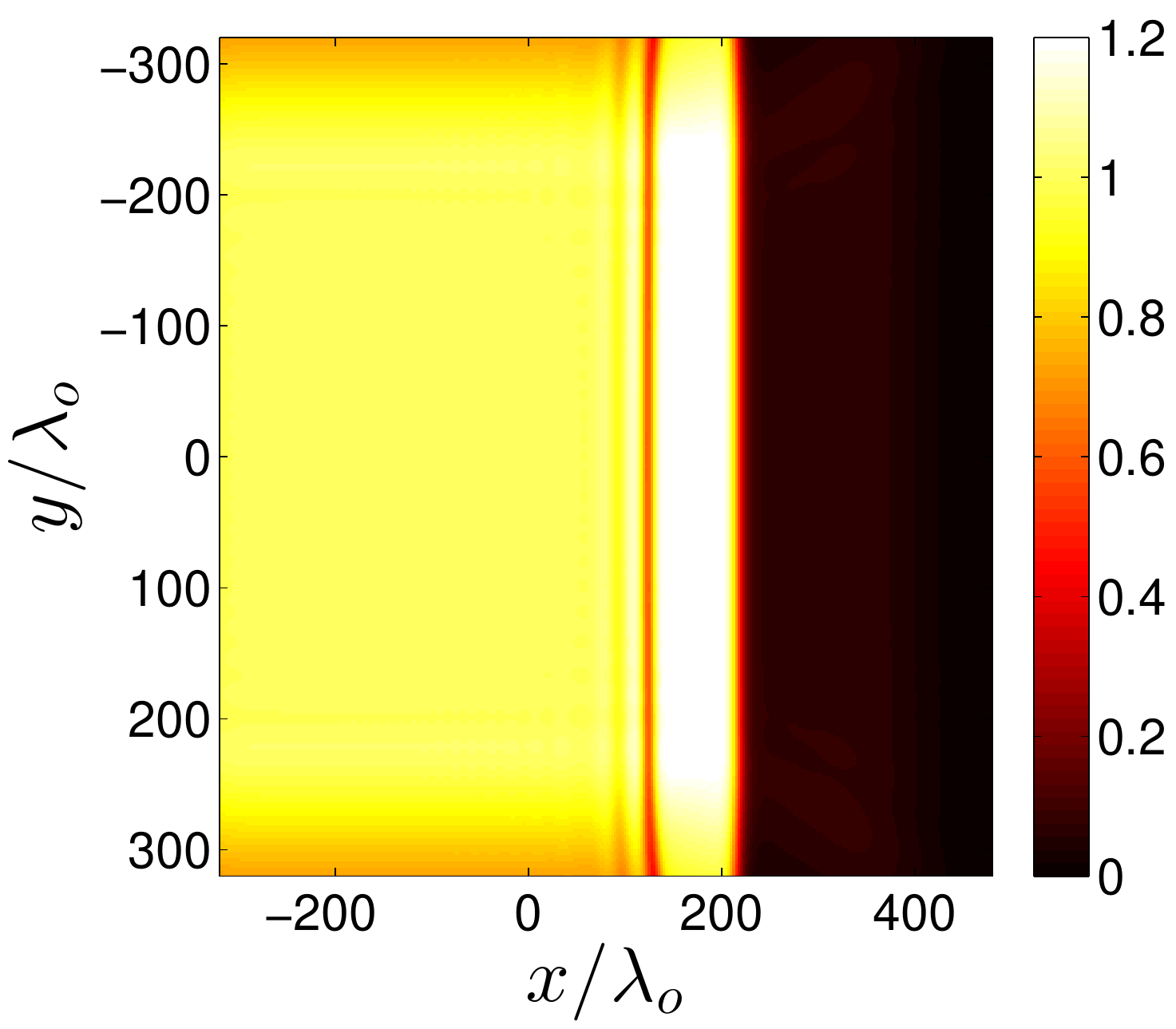}
\includegraphics[width=3cm]{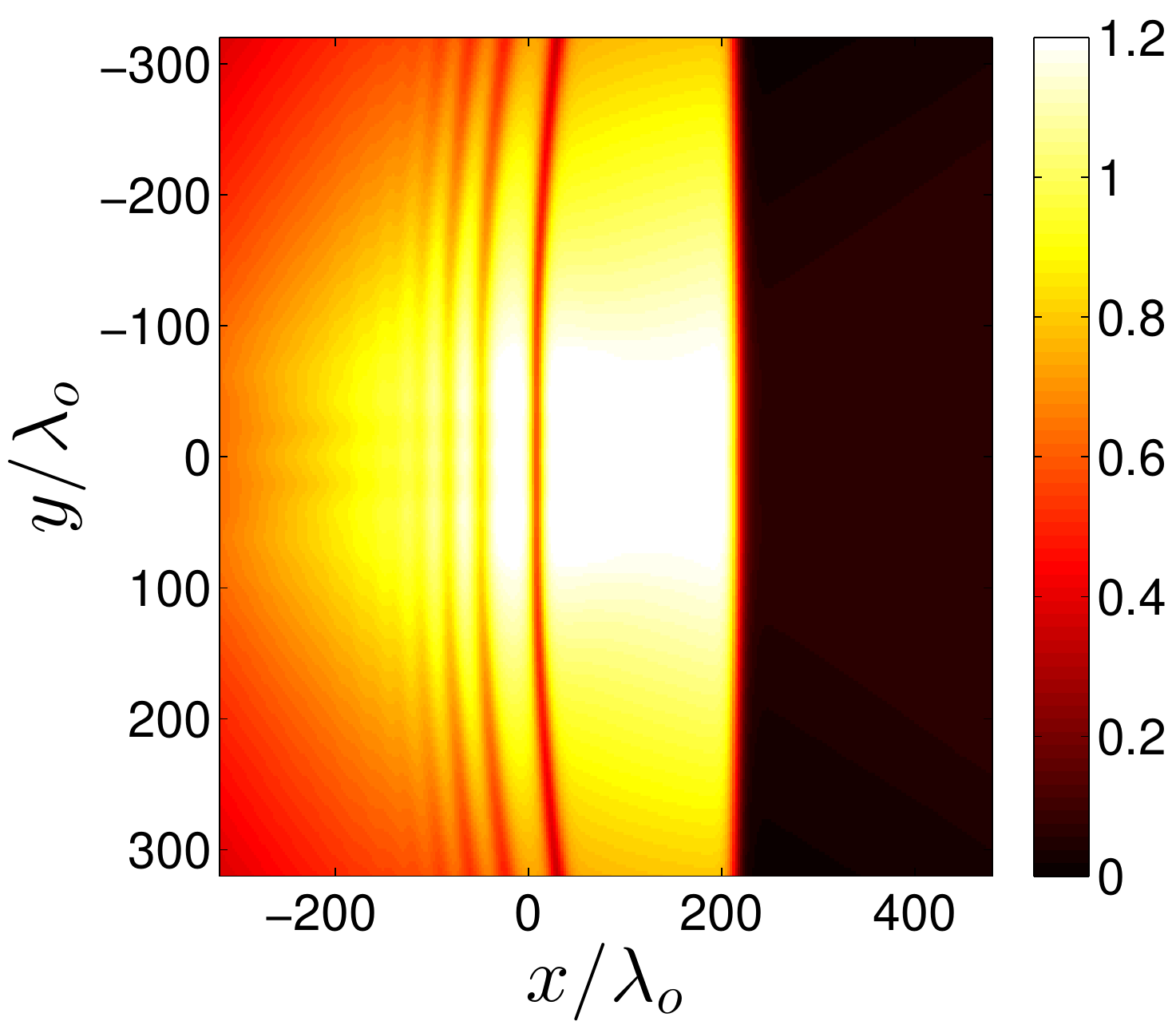}
\includegraphics[width=3cm]{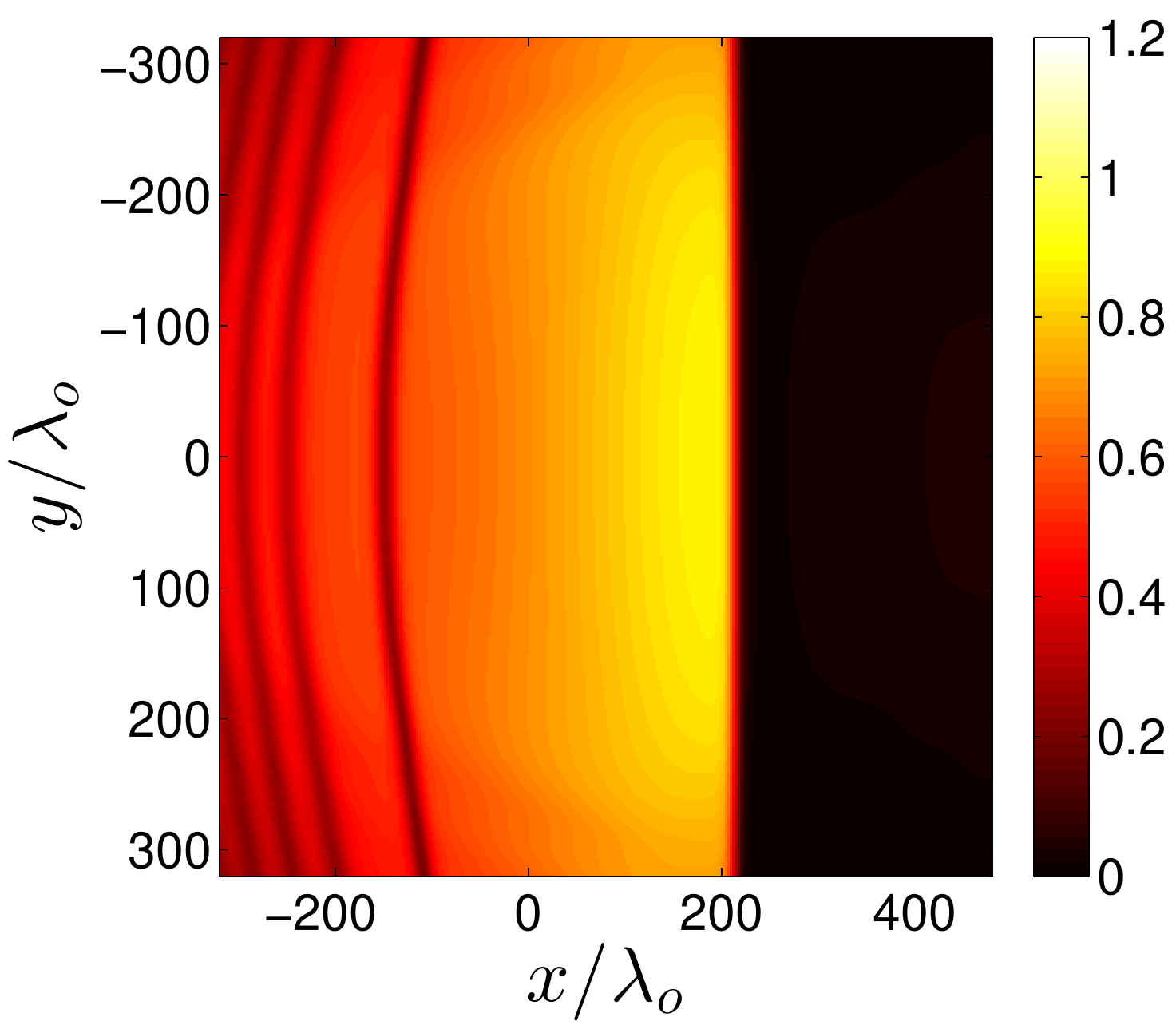} \\
\includegraphics[width=10cm]{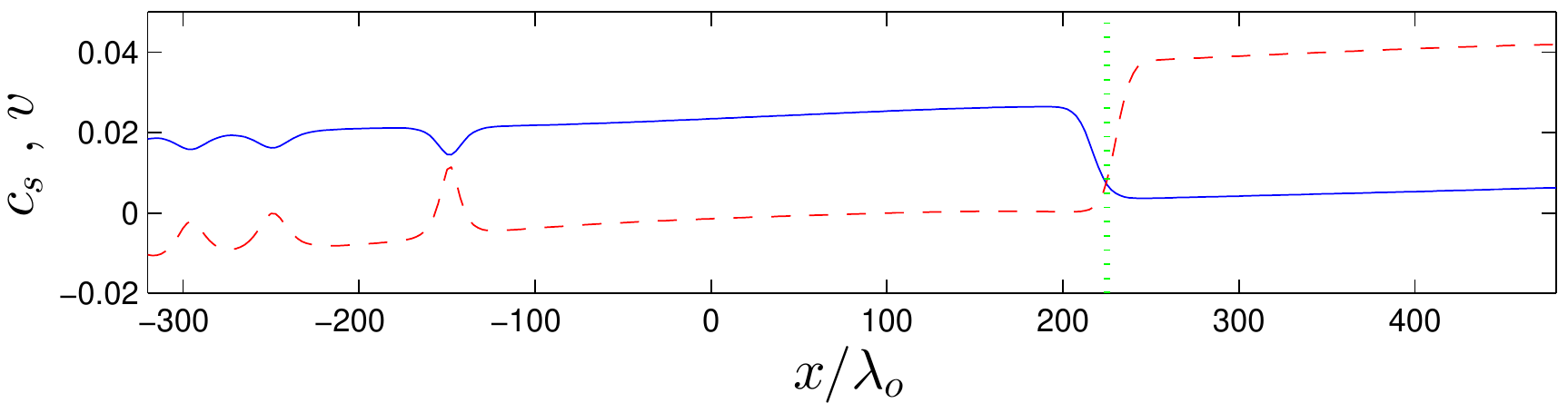}
\caption{The upper row shows a few snapshots of the tranverse intensity profile for growing propagation distances from left to right: $z/\lambda_0=0$ (incident beam, left panel), $5000$, $11000$, and $19000$ (right-most panel). The defect has a barrier shape, centered at $x_b/\lambda_0=225$ with $\delta \epsilon_{\rm max}/\epsilon=-0.0016$ and $\sigma/\lambda_0=8$. The incident beam has $v/c_s^0=0.31$. In all panels, the color scale is normalized to the incident intensity.
The lower panel shows a cut along the $y=0$ line of the local sound speed $c_s$ (solid blue line) and of the $x$ component of the local flow speed (dashed red line) at the longest propagation distance $z/\lambda_0=19000$. The crossing point of the two curves in the vicinity of the barrier (vertical dotted green line) indicates the position of the horizon.
\label{fig:bh}}
\end{center}
 \end{figure}

Fig.\ref{fig:bh} illustrates how such configurations can be realized using a barrier-shaped defect with
\begin{equation}
 \delta \epsilon(\rr_\perp,z)=\delta \epsilon_{\rm max} \, \exp (-(x-x_b)^2/2\sigma^2):
\end{equation}
the refractive index modulation of the barrier extends for the whole size of the crystal both along the propagation direction $z$ and also transversally along the $y$ direction. Along $x$ the barrier has instead a finite thickness $\sigma$. We consider a top-hat light beam focussed on one side of the barrier with a small wavevector $\kk_\perp^{\rm inc}$ directed towards it. Its intensity profile is shown in the left panel: the quite unusual square flat-top shape was chosen in the calculations to suppress all spurious features that would disturb the physics of interest.

As expected from the simulations for atomic condensates presented in \cite{pavloff}, when the light fluid hits the barrier it is partially reflected creating a series of planar fringes in front of it. During propagation along the crystal, the planar fringes form a dispersive shock wave, which is quickly expelled away from the barrier in the backward direction (central panels), leaving a clear black hole horizon in the barrier region (right panel). The trans-sonic nature of the interface is visible in the cut displayed in the bottom panel, which confirms the presence of the black hole horizon.

Following our previous work on analog black holes in atomic condensates~\cite{NJPIC,PRABH,PE}, we are presently addressing the hydrodynamic and quantum hydrodynamic properties of the horizon: scattering of phonon waves off the horizon would provide a classical counterpart of the Hawking radiation, while correlations in the transmitted light would give evidence of the true Hawking emission originating from the conversion of zero-point fluctuations into observable phonons by the horizon.

\section{Conclusions}
\label{sec:conclu}

In this article we have reviewed how the concept of fluid of light can be used to shine new light on the classical problem of the paraxial propagation of a monochromatic laser beam in a Kerr nonlinear medium. Manifestations of superfluidity such as a suppressed scattering from defects in the medium are illustrated, as well as the generation of topological excitations such as solitons and defects. The perspectives of the propagating geometry for studies of analog models of gravitational physics using fluids of light are finally outlined.

\section{Acknowledgments}
The research presented in this work builds on the long experience on superfluid light in planar microcavities that was accumulated during the years in collaboration with my many coauthors. The more speculative study of trans-sonic flows in fluids of light has strongly benefitted from discussions with Stefano Finazzi, Daniele Faccio, Pierre-\'Elie Larr\'e and Nicolas Pavloff.


\begin{thebibliography}{99}


\bibitem{ICCCRMP} I. Carusotto and C. Ciuti, Rev. Mod. Phys. {\bf 85}, 299
 (2013).

\bibitem{Amo2009} A. Amo, J. Lefr\`ere, S. Pigeon, C. Adrados, C. Ciuti, I. Carusotto, R. Houdr\'e, E. Giacobino, and A. Bramati, Nature Physics {\bf 5}, 805 (2009).

\bibitem{Amo11} A. Amo, S. Pigeon, D. Sanvitto, V. G. Sala, R. Hivet, I. Carusotto, F. Pisanello, G. Lem\'enager,  R. Houdr\'e, E. Giacobino, C. Ciuti, A. Bramati, Science {\bf 332}, 6034 (2011).

\bibitem{Nardin} G. Nardin, G. Grosso, Y. Leger, B. Pietka, F. Morier-Genoud, B. Deveaud-Pledran, Nat. Phys. {\bf 7}, 635-641 (2011).

\bibitem{Sanvitto} D. Sanvitto, S. Pigeon, A. Amo, D. Ballarini, M. De Giorgi, I. Carusotto, R. Hivet, F. Pisanello, V. G. Sala, P. S. S. Guimaraes, R. Houdr\'e, E. Giacobino, C. Ciuti, A. Bramati, G. Gigli, Nat. Phot. {\bf 5}, 610-614 (2011).

\bibitem{Wan} W. Wan, S. Jia, J. W. Fleischer, Nat. Phys. {\bf 3}, 46-51 (2007).

\bibitem{Wan2} W. Wan, S. Muenzel, J. W. Fleischer, Phys. Rev. Lett. {\bf 104}, 073903 (2010).

\bibitem{Jia} S. Jia, W. Wan, and J. W. Fleischer, Phys. Rev. Lett. {\bf 99}, 223901 (2007).

\bibitem{Wan3} W. Wan, D. V. Dylov, C.  Barsi, and J. W. Fleischer, Opt. Lett. {\bf 35}, 2819-2821 (2010).


\bibitem{Jia2} S. Jia, M. Haataja, J. W. Fleischer, New J. Phys. {\bf 7}, 075009 (2012).

\bibitem{Khamis} E. G. Khamis, A. Gammal, G. A. El, Yu. G. Gladush, A. M. Kamchatnov, Phys. Rev. A {\bf 78}, 013829 (2008).


\bibitem{Moulieras} P. Leboeuf and S. Moulieras, Phys. Rev. Lett. {\bf 105}, 163904 (2010).



\bibitem{PSbook} L.P. Pitaevskii and S. Stringari, {\em Bose-Einstein
Condensation}, Clarendon Press Oxford (2003).

\bibitem{Lukin} T. Peyronel {\em et al.}, Nature 488, 57-60 (2012).

\bibitem{ICCCPRL2004} I. Carusotto and C. Ciuti, Phys. Rev. Lett. {\bf 93}, 166401 (2004).

\bibitem{ICatoms} I. Carusotto, S. X. Hu, L. A. Collins, and A. Smerzi, Phys. Rev. Lett. {\bf 97}, 260403 (2006).

\bibitem{jena} S. Nolte, M. Will, J. Burghoff, A. Tuennermann, Appl. Phys. A, {\bf 77}, 109 (2003).

\bibitem{duck} I. Carusotto and G. Rousseaux, in Proceedings of the IX SIGRAV School on 'Analogue Gravity', Como (Italy), May 2011. Preprint available at arXiv:1202.3494.

\bibitem{opticsvortex} M. Vaupel, K. Staliunas, and C. O. Weiss, Phys. Rev. A {\bf 54}, 880 (1996)


\bibitem{Fouxon} I. Fouxon, O. V. Farberovich, S. Bar-Ad and V. Fleurov, Europhys. Lett. {\bf 92} 14002 (2010).

\bibitem{BarAd} M. Elazar, V. Fleurov, and S. Bar-Ad, Phys. Rev. A {\bf 86}, 063821 (2012).

\bibitem{Barcelo} C. Barcel\'o, S. Liberati, and M. Visser, Living Rev. Relativity {\bf 8}, 12 (2005). Available at URL: \texttt{http://www.livingreviews.org/lrr-2005-12}


\bibitem{pavloff} A. M. Kamchatnov and N. Pavloff, Phys. Rev. A {\bf 85}, 033603 (2012)


\bibitem{NJPIC} I. Carusotto, S. Fagnocchi,  A. Recati, R. Balbinot, and A. Fabbri, New J. Phys. {\bf 10}, 103001 (2008).

\bibitem{PRABH} A. Recati, N. Pavloff, and I. Carusotto, Phys. Rev. A {\bf 80}, 043603 (2009).

\bibitem{PE} P.-ƒ. LarrŽ et al., Phys. Rev. A 85, 013621 (2012)

\end{thebibliography}
\end{document}